\documentclass{scrartcl}

\usepackage[british]{babel}
\usepackage[utf8]{inputenc} 
\usepackage[T1]{fontenc}

\usepackage{mathpazo}
\usepackage{tgheros}

\usepackage{amssymb}
\usepackage{amsmath}
\usepackage{graphicx}
\usepackage{xspace}
\usepackage{upgreek}
\usepackage{csquotes}
\usepackage{booktabs}
\usepackage{multirow}
\usepackage{framed}
\usepackage{mathpartir}
\usepackage{tabularx}
\usepackage{longtable}
\usepackage{subfig}
\usepackage[page]{appendix}

\usepackage{url}
\usepackage{hyperref}

\newcommand{\pdfvalpair}[2]{\ensuremath{<\hskip-1mm|\hskip0.2mm#1,#2\hskip0.3mm|\hskip-1mm>}}
\newcommand{\pdfpair}[2]{\ensuremath{<\hskip-1mm#1,#2\hskip-1.3mm>}}
\newcommand{\pdfop}[1]{\ensuremath{#1\mathbin{\$}}}
\newcommand{\pdfcpair}[2]{\ensuremath{<\hskip-1mm#1,#2\hskip-1.3mm>_\mathrm{c}}}
\newcommand{\pdfcop}[1]{\ensuremath{#1\mathbin{\$_\mathrm{c}{}}}}
\newcommand{\pdfint}[2]{\ensuremath{{\textstyle\int_\mathrm{c}}\hskip0.2mm#1\hskip0.4mm\partial#2}}
\newcommand{\turnstile}{\vdash}
\newcommand{\exprtyping}[3]{\ensuremath{#1\turnstile #2\,:\,#3}}
\newcommand{\cexprtyping}[3]{\ensuremath{#1\turnstile_\mathrm{c} #2\,:\,#3}}
\newcommand{\exprcompd}[3]{\ensuremath{#1\turnstile_\mathrm{d} #2\,\Rightarrow#3}}
\newcommand{\exprcompc}[3]{\ensuremath{#1\turnstile_\mathrm{c} #2\,\Rightarrow#3}}
\newcommand{\exprcomp}[3]{\ensuremath{#1: #2\Rightarrow_\mathrm{c} #3}}

\newcommand{\BB}{\mathbb{B}}
\newcommand{\RR}{\mathbb{R}}
\newcommand{\ZZ}{\mathbb{Z}}

\newcommand{\kernel}{\mathbb{S}}
\newcommand{\posint}{\int^+\hskip-3.5mm}
\newcommand{\catname}[1]{\ensuremath{\mathfrak{#1}}}
\newcommand{\dd}{\partial}

\begin{document}

\title{A Verified Compiler for\\ Probability Density Functions}

%
%
\author{\Large Manuel Eberl, Johannes H\"olzl, and Tobias Nipkow\\ \large Technische Universit\"at M\"unchen}
\date{}

%
%

\newcommand{\isabellehol}{Isabelle\slash HOL\xspace}
\newcommand{\etAl}{~et~al.\xspace}
\newcommand{\ie}{i.\,e.\xspace}
\newcommand{\eg}{e.\,g.\xspace}
\newcommand{\wrt}{w.\,r.\,t.\xspace}
\newcommand{\wuppdi}[0]{\hfill\ensuremath{\square}}
\newcommand{\bind}{\gg\!\!=}
\newcommand{\hateq}{\mathrel{\hat{=}}}
\newcommand{\dbshift}{\bullet}

\newcommand\blfootnote[1]{%
  \begingroup
  \renewcommand\thefootnote{}\footnotetext{#1}%
  \addtocounter{footnote}{-1}%
  \endgroup
}

\maketitle
\blfootnote{The final publication is available at Springer via \url{https://doi.org/10.1007/978-3-662-46669-8_4}.}
\vspace*{-3em}

\begin{abstract}
Bhat\etAl~\cite{bhat13} developed an inductive compiler that computes density functions for probability spaces 
described by programs in a simple probabilistic functional language. In this work, we implement such a compiler for a 
modified version of this language within the theorem prover Isabelle and give a formal proof of its soundness \wrt 
the semantics of the source and target language. Together with Isabelle's code generation for inductive predicates, 
this yields a fully verified, executable density compiler. The proof is done in two steps, using a standard refinement approach: first, an abstract compiler working with abstract functions modelled directly in the theorem prover's logic is defined and proven sound. Then, this compiler is refined to a concrete version that returns a target-language expression.

\end{abstract}

\section{Introduction}

\subsection{Motivation}

Random distributions of practical significance can often be expressed as probabilistic functional programs. When studying a random distribution, it is often desirable to determine its \emph{probability density function} (PDF). This can be used to \eg determine the expectation or sample the distribution with a sampling method such as \emph{Markov-chain Monte Carlo} (MCMC).

In 2013, Bhat\etAl presented a compiler that computes the probability distribution function of a program in the probabilistic 
functional language \emph{Fun}~\cite{bhat13}. They evaluated the compiler on a number of practical problems and concluded that 
it reduces the amount of time and effort required to model them in an MCMC system significantly compared to hand-written models.

Bhat\etAl also stated that their eventual goal is the formal verification of such a compiler in a theorem prover \cite{bhat12}. This has the advantage of providing \emph{guaranteed correctness}, \ie the result of the compilation is provably a PDF for the source expression, according to the formal semantics. This greatly increases the confidence one has in the compiler.

In this Master's thesis, we implemented such a compiler for a similar probabilistic functional language in the interactive theorem prover \emph{Isabelle\slash HOL} and formally proved its correctness.

\subsection{Related work}

This work is based on the work of Bhat\etAl~\cite{bhat12, bhat13}, which presents a density compiler for probability spaces described by expressions in the language \emph{Fun}. This is a small functional language with basic arithmetic, Boolean logic, product and sum types, conditionals, and a number of built-in discrete and continuous distributions. It does, however, not support lists or recursion. The correctness proof is purely pen and paper, but the authors stated that formalisation in a proof assistant such as Coq is the ultimate goal~\cite{bhat12}.

Park\etAl~\cite{park05} developed a probabilistic extension to Objective CAML called $\uplambda_\bigcirc$. Their work focuses on \emph{sample generation}, \ie using an infinite stream of random numbers to compute samples of the distribution described by a probabilistic program. While Bhat\etAl generate density functions of functional programs, Park\etAl generate \emph{sampling functions}. These are functions that map $[0;1]^\infty$ to the sample space. The sampling function effectively uses a finite number of uniformly-distributed random numbers from the interval $[0;1]$ and returns a sample of the desired random variable. This approach allows them to handle much more general distributions, even recursively-defined ones and distributions that do not have a density function, but it does not allow precise reasoning about these distributions (such as determining the exact expectation). No attempt at formal verification is made.

pGCL, another probabilistic language, \emph{has} been formalised with a proof assistant -- first in HOL by Hurd\etAl~\cite{hurd}, then in Isabelle\slash HOL by David Cock~\cite{cock, cock_isa}. pGCL contains a rich set of language features, such as recursion and probabilistic and non-deterministic choice. David Cock formally proved a large number of results about the semantics of the language and developed mechanisms for refinement and program verification; however, the focus of this work was verification of probabilistic programs, not compiling them to density functions. Bhat\etAl mention that reconciling recursion with probability density functions is difficult~\cite{bhat12}, so a feature-rich language such as pGCL is probably not suited for developing a density compiler.

Audebaud and Paulin-Mohring~\cite{audebaud} also implemented a probabilistic functional language with recursion in a theorem prover (namely Coq). Their focus was also on verification of probabilistic programs. While Cock uses a shallow embedding in which any type and operation of the surrounding logic of the theorem prover (\ie HOL) can be used, Audebaud and Paulin-Mohring use a deep embedding with a restricted type system, expression structure, predefined primitive functions, etc. Like Bhat\etAl and we, they also explicitly reference the \emph{Giry monad} as a basis for their work.

\subsection{Utilised tools}

As stated before, we work with the interactive theorem prover Isabelle\slash HOL. Isabelle is a generic proof assistant that supports a number of logical frameworks (object logics) such as \emph{Higher-Order Logic} (HOL) or \emph{Zermelo-Fraenkel set theory} (ZF). The most widely used instance is Isabelle\slash HOL, which is what we use.

We heavily rely on Johannes H\"olzl's Isabelle formalisation of measure theory~\cite{hoelzl}, which is already part of the Isabelle\slash HOL library. We also use Sudeep Kanav's formalisation of the Gaussian distribution and a number of libraries from the proof of the Central Limit Theorem by Avigad\etAl~\cite{avigad}, namely the notion of interval and set integrals and the Fundamental Theorem of Calculus for these integrals. All of these have been or will be moved to the Isabelle\slash HOL library by their respective maintainers as well.

\subsection{Outline}

In Sect. \ref{sec_notation}, we will explain the notation we will use and then give a brief overview of the mathematical basics we require -- in particular the Giry Monad -- in Sect. \ref{sec_mathematical}.

Section \ref{sec_language} contains the definition and the semantics of the source and target language. Section \ref{sec_abstract} defines the abstract compiler and gives a high-level outline of the soundness proof. Section \ref{sec_concrete} then explains the refinement of the abstract compiler to the concrete compiler and the final correctness result and evaluates the compiler on a simple example.

Section \ref{sec_conclusion} gives an overview of how much work went into the different parts of the project and what difficulties we encountered. It also lists possible improvements and other future work before summarising the results we obtained. Finally, the appendix contains a table of all notation and auxiliary functions used in this work for reference.

\section{Preliminaries}
\label{sec_preliminaries}

\subsection{Notation}
\label{sec_notation}

In the following, we will explain the conventions and the notation we will use.

\subsubsection{Typographical notes}

We will use the following typographical conventions in mathematical formul\ae:
\begin{itemize}
\item Constants, functions, datatype constructors, and types will be typeset in regular roman font: $\mathrm{max}$, $\mathrm{return}$, $\uppi$, etc.\footnote{In formul\ae\ embedded in regular text, we will often set types and constants in italics to distinguish them from the surrounding text.}
\item Free and bound variables (including type variables) are in italics: $x$, $A$, $\mathit{dst}$, etc.
\item Isabelle keywords are in bold print: $\mathbf{lemma}$, $\mathbf{datatype}$, $\mathbf{primrec}$, etc.
\item $\upsigma$-algebras are typeset in calligraphic font: $\mathcal A$, $\mathcal B$, $\mathcal M$, etc.
\item Categories are typeset in Fraktur: $\mathfrak{Set}$, $\mathfrak{Grp}$, $\mathfrak{Meas}$, etc.
\item File names of Isabelle theories are set in a monospaced font: \texttt{PDF\_Compiler.thy}.
\end{itemize}

\subsubsection{Deviations from standard mathematical notation}

In order to maintain coherence with the notation used in Isabelle, we will deviate from standard mathematical notation 
in the following ways:
\begin{itemize}
\item As is convention in functional programming (and therefore in Isabelle), function application is often written as 
$f\ x$ instead of $f(x)\;$. It is the operation that binds strongest and it associates to the left, \ie $f\ x + 1$ is $(f\ x) + 1$ and $f\ x\ y$ is $(f\ x)\ y\;$.
\item The integral over integrand $f$ with variable $x$ and the measure $\mu$ is written as
$$\int\hskip-1mm x.\ f\ x\ \dd\mu\hskip5mm\mathrm{instead\ of}\hskip5mm\int\hskip-0.5mm f(x)\,\mathrm{d}\mu(x)\;.$$
\item The special notion of a \emph{non-negative integral}\footnote{This is a very central concept in the measure theory library in Isabelle. We will mostly use it with non-negative functions anyway, so the distinction is purely formal.} is used; this integral \enquote{ignores} the negative part of the integrand. Effectively:
$$\posint x.\ f\ x\ \dd\mu\hskip5mm\hateq\hskip5mm\int\hskip-0.5mm \mathrm{max}(0, f(x))\, \mathrm{d}\mu(x)$$
\item Lambda abstractions, such as $\uplambda x.\ x^2$, are used to specify functions. The corresponding standard 
mathematical notation would be $x\mapsto x^2\;$.
\end{itemize}

Table \ref{tbl:notation} lists a number of additional mathematical notation.

\renewcommand{\arraystretch}{1.2}
\begin{table}[bpt]
\caption{General notation}
\label{tbl:notation}
{\small
\begin{tabularx}{\textwidth}{lllX}
\toprule
\textsc{Notation} & \textsc{Description} & \hspace*{2mm} & \textsc{Definition} \\\midrule
$f\ `\,X$ & image set & & $f(X)$ or $\{f(x)\ |\ x\in X\}$\\
$\mathrm{undefined}$ & arbitrary value & & \\
$\mathrm{merge}\ V\ V'\ (\rho, \sigma) \hspace*{2mm}$ & merging disjoint states & & $\begin{cases}\rho\ x & \mathrm{if}\ x\in V\\\sigma\ y & \mathrm{if}\ x\in V'\\\mathrm{undefined} & \mathrm{otherwise}\end{cases}$\\[2mm]
$\langle P\rangle$ & indicator function & & $1$ if $P$ is true, $0$ otherwise\\
$\mathrm{density}\ M\ f$ & measure with density & & result of applying density $f$ to $M$\\
$\mathrm{distr}\ M\ N\ f$ & push-forward/image measure & & $(B,\ \mathcal B,\ \uplambda X.\ \mu(f^{-1}(X)))$ for\\
& & & $M = (A,\mathcal A, \mu)$, $N = (B, \mathcal B, \mu')$\\
\bottomrule
\end{tabularx}}
\end{table}
\renewcommand{\arraystretch}{1.0}

\subsubsection{Semantics notation}
\label{sec_notation_semantics}

We will always use $\Gamma$ to denote a type environment, \ie a function from variable names to types, and $\sigma$ to denote a state, \ie a function from variable names to values. Note that variable names are always natural numbers as we use de~Bruijn indices. The letter $\Upsilon$ (Upsilon) will be used to denote \emph{density contexts}, which are used in the compiler.

We also employ the notation $t\dbshift\Gamma$ resp. $v\dbshift\sigma$ to denote the insertion of a new variable with the type $t$ (resp. value $v$) into a typing environment (resp. state). This is used when entering the scope of a bound variable; the newly inserted variable then has index 0 and all other variables are incremented by 1. The precise definition is as follows:\footnote{
Note the analogy to the notation $x\mathbin{\#}xs$ for prepending an element to a list. This is because contexts with de~Bruijn indices are 
generally represented by lists, where the $n$-th element is the entry for the variable $n$, and inserting a value for a newly bound 
variable is then simply the prepending operation.}

$$(y\dbshift f)(x) = \begin{cases}y & \mathrm{if}\ x = 0\\f(x-1) & \mathrm{otherwise}\end{cases}$$

We use the same notation for inserting a new variable into a set of variables, shifting all other variables, \ie:

$$x\dbshift V = \{x\} \cup \{y+1\ |\ y\in V\}$$

In general, the notation $\exprtyping{\Gamma}{e}{t}$ and variations thereof will always mean \enquote{The expression $e$ has type $t$ in the type environment $\Gamma$}, whereas $\Upsilon\turnstile e \Rightarrow f$ and variations thereof mean \enquote{The expression $e$ compiles to $f$ under the context $\Upsilon$}.

\subsection{Mathematical basics}
\label{sec_mathematical}

The category theory part of this section is based mainly on a presentation by Ernst-Erich Doberkat~\cite{doberkat}. 
For a more detailed introduction, see his textbook~\cite{doberkat_book} or the original paper by Mich\`ele Giry~\cite{giry}.

\subsubsection{Basic measure spaces.} There are two basic measure spaces we will use:
\paragraph{Counting space.} The counting space on a countable set $A$ has
\begin{itemize}
\item the carrier set $A$
\item the measurable sets $\mathcal P(A)$, \ie all subsets of $A$
\item the measure $\uplambda X.\ |X|$, \ie the measure of a subset of $A$ is simply the number of elements in that set (which can be $\infty$)
\end{itemize}
\paragraph{Borel space.} The Borel space is a measure space on the real numbers which has
\begin{itemize}
\item the carrier set $\RR$
\item the Borel $\upsigma$-algebra as measurable sets, \ie the smallest $\upsigma$-algebra containing all open subsets of $\RR$
\item the Borel measure as a measure, \ie the uniquely defined measure that maps real intervals to their lengths
\end{itemize}

\subsubsection{Sub-probability spaces.}\label{subprob} A sub-probability space is a measurable space $(A, \mathcal A)$ with a measure  $\mu$ such that every set $X\in \mathcal A$ has a measure $\leq 1$, or, equivalently, $\mu(A)\leq 1$\;.

For technical reasons, we also assume $A\neq\emptyset$\;. This is required later in order to define the $\mathit{bind}$ operation in the Giry monad in a convenient way within Isabelle. 
This non-emptiness condition will always be trivially satisfied by all the measure spaces used in this work.

\subsubsection{The category $\catname{Meas}$.}\label{cattheory} Note that:
\begin{itemize}
\item For any measurable space $(A,\mathcal A)$, the identity function is $\mathcal A$-$\mathcal A$-measurable.
\item For any measurable spaces $(A,\mathcal A)$, $(B, \mathcal B)$, $(C, \mathcal C)$, an $\mathcal A$-$\mathcal B$-measurable function $f$, 
and a $\mathcal B$-$\mathcal C$-measurable function $g$, the function $g \circ f$ is $\mathcal A$-$\mathcal C$-measurable.
\end{itemize}

\noindent Therefore, measurable spaces form a category \catname{Meas} where:
\begin{itemize}
\item the objects of the category are measurable spaces
\item the morphisms of the category are measurable functions
\item the identity morphism $\mathbf 1_{(A,\mathcal A)}$ is the identity function $\mathrm{id}_A : A \to A,\ \uplambda x.\ x$
\item morphism composition is function composition
\end{itemize}

\subsubsection{Kernel space.}\label{kernelspace} The kernel space $\kernel(A, \mathcal A)$ of a measurable space $(A, \mathcal A)$ is the natural measurable 
space over the measures over $(A, \mathcal A)$ with a certain property. For our purposes, this property will be that 
they are sub-probability measures (as defined in Sect. \ref{subprob}).

Additionally, a natural property the kernel space should satisfy is that measuring a fixed set $X\in\mathcal A$ 
while varying the measure within $\kernel(A, \mathcal A)$ should be a Borel-measurable function; formally:
$$\mathrm{For\ all}\ X\in \mathcal A\mathrm{,\ }f\hskip-0.9mm: \kernel(A, \mathcal A) \to \RR,\ \uplambda \mu.\ \mu(X)\textrm{\,\ is\ \,}\mathcal \kernel(A, \mathcal A)\textrm{-Borel-measurable}$$

\noindent\begin{minipage}{\textwidth}
We can now simply define $\kernel(A, \mathcal A)$ as the smallest measurable space with the carrier set
$$M := \{\mu\ |\ \mu\ \mathrm{is\ a\ measure\ on}\ (A, \mathcal A),\ \mu(A) \leq 1\}$$
that fulfils this property, \ie we let $\kernel(A, \mathcal A) := (M, \mathcal M)$ with
$$\mathcal M := \{(\uplambda \mu.\ \mu(X))^{-1}(Y)\ |\ X\in\mathcal A, Y\in \mathcal B\}$$
where $\mathcal B$ is the Borel $\upsigma$-algebra on $\RR$.
\end{minipage}

Additionally, for a measurable function $f$, we define $\kernel(f) = \uplambda\mu.\ f(\mu)$, where $f(\mu)$ denotes the push-forward measure (or image measure)\footnote{$f(\mu) = \uplambda X.\ \mu(f^{-1}(X))$\;.}. Then $\kernel$ maps objects of \catname{Meas} to objects of \catname{Meas} and morphisms of $\catname{Meas}$ to morphisms of $\catname{Meas}$. We can thus see that $\kernel$ is an endofunctor in the category \catname{Meas}, as $(\mathrm{id}_{(A,\mathcal A)})(\mu) = \mu$ and $(f\circ g)(\mu)=f(g(\mu))$\;.

\subsubsection{Giry monad.}\label{giry} The Giry monad naturally captures the notion of choosing a value according to a \mbox{(sub-)}prob\-a\-bil\-ity distribution, using it as a parameter for another distribution, and observing the result.

Consequently, $\mathit{return}$ (or $\upeta$) yields a Dirac measure, \ie a probability measure in which all the \enquote{probability} lies in a single element, and 
$\mathit{bind}$ (or $\bind$) integrates over all the input values to compute one single output measure. Formally, for 
measurable spaces $(A, \mathcal A)$ and $(B, \mathcal B)$, a measure $\mu$ on $(A, \mathcal A)$, a value $x\in A$, and an $\mathcal A$-$\kernel(B, \mathcal B)$-measurable 
function $f$:
$$\mathrm{return}_{(A, \mathcal A)}\ x := \uplambda X.\ \begin{cases}1&\mathrm{if}\ x\in X\\0&\mathrm{otherwise}\end{cases}\hskip5mm
\mu \bind f := \uplambda X.\ \int\hskip-1mm x.\ f(x)(X) \,\dd\mu$$

Unfortunately, restrictions due to Isabelle's type system require us to determine the $\upsigma$-algebra of the resulting measurable space for $\mathit{bind}\ M\ f$ since this information cannot be provided by the type of $f$. This can be done with an additional parameter, but it is more convenient to define bind in such a way that it chooses an arbitrary value $x\in M$ and takes the $\upsigma$-algebra of $f\ x$ (or the count space on $\emptyset$ if $M$ is empty)\footnote{Note that for any $\mathcal{A}-\kernel(B,\mathcal B)$-measurable function $f$, the $\upsigma$-algebra thus obtained is independent of which value is chosen, by definition of the kernel space.}. 

This choice is somewhat non-standard, but the difference is of no practical significance as we will not use $\mathit{bind}$ on empty measure spaces.

To simplify the proofs for \emph{bind}, we instead define the \emph{join} operation (also known as $\upmu$ in category theory) first and use it to then define \emph{bind}. The \emph{join} operation \enquote{flattens} objects, \ie it maps 
an element of $\kernel(\kernel(A, \mathcal A))$ to one of $\kernel(A, \mathcal A)$\;. Such an operation can be naturally defined as:
$$\mathrm{join}\ \mu = \uplambda X.\ \int\hskip-1mm \mu'.\ \mu'(X) \,\dd\mu$$

Note that in Isabelle, \emph{join} has an additional explicit parameter for the measurable space of the result to avoid the problem we had with \emph{bind}. This makes expressions containing \emph{join} more complicated; this is, however, justified by the easier proofs and will be unproblematic later since we will never use \emph{join} directly, only \emph{bind}.

Now, \emph{bind} can be defined using \emph{join} in the following way, modulo handling of empty measure spaces\footnote{$\kernel(f)$, \ie lifting a function on values to a function on measure spaces, is done by the \emph{distr} function in Isabelle. This function was already defined in the library, which simplifies our proofs about \emph{bind}, seeing as some of the proof work has already been done in the proofs about \emph{distr}.}:
$$\mu \bind f = \mathrm{join}\ (\kernel(f)(\mu)) = \mathrm{join}\ (f(\mu))$$

\label{monadsyntax}
\subsubsection{The \enquote{do} syntax.} For better readability, we employ a Haskell-style \enquote{\textbf{do} notation} for operations in the Giry monad. The syntax of this notation 
is defined recursively, where $M$ stands for a monadic expression and $\langle\hskip0.2mm\mathit{pattern}\rangle$ stands for arbitrary \enquote{raw} text:
$$\mathbf{do}\ \{M\} \,\hateq\, M\hskip7mm 
\mathbf{do}\ \{x \leftarrow M;\ \mathrm{\langle\hskip0.2mm  \mathit{pattern}\rangle}\} \,\hateq\, M \bind (\uplambda x.\ \mathbf{do}\ \{\langle \mathit{\hskip0.2mm pattern}\rangle\})$$

\section{Language Syntax and Semantics}
\label{sec_language}

The source language used in the formalisation was modelled after the language \emph{Fun} described by Bhat\etAl~\cite{bhat13}; similarly, the target language is almost identical to the target language used by Bhat\etAl However, we have made the following changes in our languages:
\begin{itemize}
\item Variables are represented by de~Bruijn indices to avoid handling freshness, capture-avoiding substitution, and related problems.
\item No sum types are supported. Consequently, the $\textbf{match}\,\ldots\,\textbf{with}\,\ldots$ command is replaced with an $\textbf{IF}\,\ldots\,\textbf{THEN}\,\ldots\,\textbf{ELSE}\,\ldots$ command. Furthermore, booleans are a primitive type rather than represented as $\mathit{unit} + \mathit{unit}$\;.
\item The type $\mathit{double}$ is called $\mathit{real}$ and it represents a real number with absolute precision as opposed to an IEEE $754$ floating point number.
\item Beta and Gamma distributions are not included.
\end{itemize}

In the following sections, we give the precise syntax, typing rules, and semantics of both our source language and our target language.\\

\subsection{Types, values, and operators}
\label{sec_types_values}

The source language and the target language share the same type system and the same operators. Figure \ref{types_values} shows the types and values that exist in our languages.\footnote{Note that $\mathit{bool}$, $\mathit{int}$, and $\mathit{real}$ stand for the respective Isabelle types, whereas $\BB$, $\ZZ$, and $\RR$ stand for the source-/target-language types.} Additionally, standard arithmetical and logical operators exist.

\begin{figure}
\begin{oframed}
\vspace*{-6mm}
\begin{align*}
&\mathbf{datatype}\ \mathrm{pdf\hskip-0.3mm\_type}\ =\ \\
&\hskip10mm\mathrm{UNIT}\ |\ \BB\ |\ \ZZ\ |\ \RR\ |\ \mathit{pdf\hskip-0.5mm\_type}\times\mathit{pdf\hskip-0.5mm\_type}\\
&\mathbf{datatype}\ \mathrm{val}\ =\ \\
&\hskip10mm\mathrm{UnitVal}\ |\ \mathrm{BoolVal}\ \mathit{bool}\ |\ \mathrm{IntVal}\ \mathit{int}\ |\ \mathrm{RealVal}\ \mathit{real}\ |\ \pdfvalpair{\mathit{val}}{\mathit{val}}\\
&\mathbf{datatype}\ \mathrm{pdf\hskip-0.3mm\_operator}\ =\ \\
&\hskip10mm\mathrm{Fst}\ |\ \mathrm{Snd}\ |\ \mathrm{Add}\ |\ \mathrm{Mult}\ |\ \mathrm{Minus}\ |\ \mathrm{Less}\ |\ \mathrm{Equals}\ |\ \mathrm{And}\ |\ \mathrm{Or}\ |\ \mathrm{Not}\ |\ \mathrm{Pow}\ |\ \\
&\hskip10mm\mathrm{Fact}\ |\ \mathrm{Sqrt}\ |\ \mathrm{Exp}\ |\ \mathrm{Ln}\ |\ \mathrm{Inverse}\ |\ \mathrm{Pi}\ |\ \mathrm{Cast}\ \mathit{pdf\hskip-0.5mm\_type}
\end{align*}
\vspace*{-8mm}
\end{oframed}
\caption{The types and values in the source and target language}
\label{types_values}
\end{figure}

All operators are \emph{total}, meaning that for every input value of their parameter type, they return a single value of their result type. This requires some non-standard definitions for non-total operations such as division, the logarithm, and the square root -- we define them to be 0 outside their domain. Non-totality could also be handled by implementing operators in the Giry monad by letting them return either a Dirac distribution with a single result or, when evaluated for a parameter on which they are not defined, the null measure. This, however, would probably complicate many proofs significantly.

\noindent \begin{minipage}{\textwidth}
\noindent To increase readability, we will use the following abbreviations:\nopagebreak[4]
\begin{itemize}
\item $\mathit{TRUE}$ and $\mathit{FALSE}$ stand for $\mathit{BoolVal\ True}$ and $\mathit{BoolVal\ False}$, respectively.
\item $\mathit{RealVal}$, $\mathit{IntVal}$, etc.\ will be omitted in expressions when their presence is implicitly clear from the context.
\item $a - b$ stands for $a + (-b)$ and $a / b$ for $a \cdot b^{-1}$\;.
\end{itemize}
\end{minipage}

\subsection{Auxiliary definitions}

A number of auxiliary definitions are used in the definition of the semantics; for a full list of auxiliary functions see table \ref{tbl:auxiliary}. The following two notions require a detailed explanation:

\subsubsection{Measure embeddings.}\label{sec_embeddings} A \emph{measure embedding} is the measure space obtained by \enquote{tagging} values in a measure space $M$ with some injective function $f$ (in fact, $f$ will always be a datatype constructor). For instance, a set of values of type $\RR$ can naturally be measured by stripping away the $\mathit{RealVal}$ constructor and using a measure on real numbers (\eg the Lebesgue-Borel measure) on the resulting set of reals. Formally:
$$\mathrm{embed\_measure}\ (A,\mathcal A,\mu)\ f\ =\ \left(f(A),\ \{f(X)\ |\ X\in\mathcal{A}\},\ \uplambda X.\ \mu(f^{-1}(X) \cap A)\right)$$

\subsubsection{Stock measures.}\label{sec_stock_measures} The \emph{stock measure} for a type $t$ is the \enquote{natural} measure on values of that type. This is defined as follows:
\begin{itemize}
\item For the countable types $\mathit{UNIT}$, $\BB$, $\ZZ$: the count measure over the corresponding type universes
\item For type $\RR$: the embedding of the Lebesgue-Borel measure on $\RR$ with $\mathit{RealVal}$

\item For $t_1 \times t_2$: the embedding of the product measure
$$\mathrm{stock\_measure}\ t_1\ {\textstyle{\bigotimes}}\ \mathrm{stock\_measure}\ t_2$$
with $\uplambda(v,w).\,\pdfvalpair{v}{w}$
\end{itemize}

Note that in order to save space and increase readability, we will often write $\int\hskip-0.8mm x.\ f\ x\,\dd t$ instead of $\int\hskip-0.8mm x.\ f\ x\,\dd\hskip0.3mm \mathit{stock\_measure}\ t$ in integrals.

\subsubsection{The state measure.} Using the stock measure, we can also construct a measure on \emph{states} in the context of a typing environment $\Gamma$. A state on the variables $V$ is a function that maps a variable in $V$ to a value. A state $\sigma$ is \emph{well-formed} \wrt to $V$ and $\Gamma$ if it maps every variable $x\in V$ to a value of type $\Gamma\ x$ and every variable ${\notin}\,V$ to $\mathit{undefined}$.

We now fix $\Gamma$ and a finite $V$ and consider the set of well-formed states \wrt $V$ and $\Gamma$. Another representation of these states are tuples in which the $i$-th component is the value of the $i$-th variable in $V$. The natural measure that can be given to such tuples is then the finite product measure of the stock measures of the types of the variables:

$$\mathrm{state\_measure}\ \Gamma\ V := \bigotimes_{x\in V}\,\mathrm{stock\_measure}\ (\Gamma\ x)$$\vspace*{-10mm}

\renewcommand{\arraystretch}{1.2}
\begin{table}[!htbp]
\caption{Auxiliary functions}
\begin{center}
\small
\noindent\begin{tabularx}{\textwidth}{lX}
\toprule
\textsc{Function} & \textsc{Description} \\\midrule
$\mathrm{op\_sem}\ \mathit{op}\ \mathit{v}$ & semantics of operator $\mathit{op}$ applied to $v$\\
$\mathrm{op\_type}\ \mathit{op}\ t$ & result type of operator $\mathit{op}$ for input type $t$\\
$\mathrm{dist\_param\_type}\ \mathit{dst}$ & parameter type of the built-in distribution $\mathit{dst}$\\
$\mathrm{dist\_result\_type}\ \mathit{dst}$ & result type of the built-in distribution $\mathit{dst}$\\
$\mathrm{dist\_measure}\ \mathit{dst}\ x$ & built-in distribution $\mathit{dst}$ with parameter $x$\\
$\mathrm{dist\_dens}\ \mathit{dst}\ x\ y$ & density of the built-in distribution $\mathit{dst}$ w. parameter $x$ at value $y$\\
$\mathrm{type\_of}\ \Gamma\ e$ & the unique $t$ such that $\exprtyping{\Gamma}{e}{t}$\\
$\mathrm{val\_type}\ v$ & the type of value $v$, \eg $\mathrm{val\_type}\ (\mathrm{IntVal}\ 42) = \mathrm{INTEG}$\\
$\mathrm{type\_universe}\ t$ & the set of values of type $t$\\
$\mathrm{countable\_type}\ t$ & $\mathrm{true}$ iff $\mathrm{type\_universe}\ t$ is a countable set\\
$\mathrm{free\_vars}\ e$ & the free (\ie non-bound) variables in the expression $e$\\
$e\ \mathrm{det}$ & $\mathrm{true}$ iff $e$ does not contain $\mathrm{Random}$ or $\mathrm{Fail}$\\
$\mathrm{extract\_real}\ x$ & returns $y$ for $x = \mathrm{RealVal}\ y$ (analogous for int, pair, etc.)\\
$\mathrm{return\_val}\ v$ & $\mathrm{return}\ (\mathrm{stock\_measure}\ (\mathrm{val\_type}\ v))\ v$\\
$\mathrm{null\_measure}\ M$ & measure with same measurable space as $M$, but 0 for all sets\\
\bottomrule
\end{tabularx}
\end{center}
\label{tbl:auxiliary}
\end{table}
\renewcommand{\arraystretch}{1.0}

\subsection{Source language}

Figures \ref{src_lang} and \ref{expr_typing} show the syntax resp. the typing rules of the source language. Figure \ref{expr_sem} defines the source language semantics as a primitively recursive function.
Similarly to the abbreviations mentioned in Sect. \ref{sec_types_values}, we will omit $\mathit{Val}$ when its presence is implicitly obvious from the context; e.\,g.\ if in some context, $e$ is an expression and $c$ is a constant real number, we will write $e + \mathit{Val}\ (\mathit{RealVal}\ c)$ as $e + c$\;.\\

\subsubsection{Built-in distributions.} Our language contains the built-in distributions \emph{Bernoulli}, \emph{UniformInt}, \emph{UniformReal}, \emph{Gaussian}, and \emph{Poisson}. The uniform distributions are parametrised with a pair of numbers $(a,b)$ and return a uniform distribution over the interval $[a;b]$. The Gaussian distribution is parametrised with a pair of real numbers $(\mu,\sigma)$, \ie mean and standard deviation.

For invalid parameters, \ie $\mathit{Bernoulli}\ 2$ or $\mathit{UniformReal}\ 3\ 2$, the built-in distributions return the null measure.

\begin{figure}
\begin{oframed}
\vspace*{-6mm}
\begin{align*}
&\mathbf{datatype}\ \mathrm{expr}\ =\ \\
&\hskip10mm\mathrm{Var}\ \mathit{nat}\ |\ \mathrm{Val}\ \mathit{val}\ |\ \mathrm{LET}\ \mathit{expr}\ \mathrm{IN}\ \mathit{expr}\ |\ \pdfop{\mathit{pdf\hskip-0.4mm\_operator}} \mathit{expr}\ |\ \hskip-0.9mm\pdfpair{\mathit{expr}}{\mathit{expr}}\hskip-1mm\ |\ \\
&\hskip10mm\mathrm{Random}\ \mathit{pdf\hskip-0.4mm\_dist}\ |\ \mathrm{IF}\ \mathit{expr}\ \mathrm{THEN}\ \mathit{expr}\ \mathrm{ELSE}\ \mathit{expr}\ |\ \mathrm{Fail}\ \mathit{pdf\hskip-0.5mm\_type}
\end{align*}
\vspace*{-8mm}
\end{oframed}
\caption{The source language syntax}
\label{src_lang}
\end{figure}

\begin{figure}
\begin{oframed}
$$\inferrule[et\_val]{ }{\exprtyping{\Gamma}{\mathrm{Val}\ v}{\mathrm{val\_type}}\ v}\quad\quad\quad
\inferrule[et\_var]{ }{\exprtyping{\Gamma}{\mathrm{Var}\ x}{\Gamma\ x}}\quad\quad\quad
\inferrule[et\_fail]{ }{\exprtyping{\Gamma}{\mathrm{Fail}\ t}{t}}$$
$$\inferrule[et\_op]{\exprtyping{\Gamma}{e}{t} \\ \mathrm{op\_type}\ \mathit{op}\ t = \mathrm{Some}\ t'}
          {\exprtyping{\Gamma}{\,\pdfop{\mathit{op}} e}{t'}}\quad\quad\quad
\inferrule[et\_pair]{\exprtyping{\Gamma}{e_1}{t_1} \\ \exprtyping{\Gamma}{e_2}{t_2}}
          {\exprtyping{\Gamma}{\,\pdfpair{e_1}{e_2}\,}{t_1\hskip-0.6mm \times t_2}}$$\vskip-2mm
$$\inferrule[et\_if]{\exprtyping{\Gamma}{b}{\BB} \\ \exprtyping{\Gamma}{e_1}{t} \\ \exprtyping{\Gamma}{e_2}{t}}
          {\exprtyping{\Gamma}{\mathrm{IF}\ b\ \mathrm{THEN}\ e_1\ \mathrm{ELSE}\ e_2}{t}}\quad\quad\ 
\inferrule[et\_let]{\exprtyping{\Gamma}{e_1}{t_1}\\ \exprtyping{t_1\dbshift\Gamma}{e_2}{t_2}}
          {\exprtyping{\Gamma}{\mathrm{LET}\ e_1\ \mathrm{IN}\ e_2}{t_2}}$$
$$\inferrule[et\_rand]{\exprtyping{\Gamma}{e}{\mathrm{dist\_param\_type}\ \mathit{dst}}}
          {\exprtyping{\Gamma}{\mathrm{Random}\ \mathit{dst}\ e}{\mathrm{dist\_result\_type}\ \mathit{dst}}}$$\vskip-2mm
\end{oframed}
\caption{The typing rules for source-language expressions}
\label{expr_typing}
\end{figure}

\subsection{Deterministic expressions}
\label{sec_randomfree}

We call an expression $e$ \emph{deterministic} (written as \enquote{$e\ \mathrm{det}$}) if it contains no occurrence of $\mathit{Random}$ or $\mathit{Fail}$. Such expressions are of particular interest: if all their free variables have a fixed value, they return precisely one value, so we can define a function $\mathit{expr\_sem\_rf}$\footnote{In Isabelle, the expression \emph{randomfree} is used instead of \emph{deterministic}, hence the \enquote{rf} suffix. This is in order to emphasise the syntactical nature of the property. Additionally, it is worth noting that a syntactically deterministic expression is not truly deterministic if the variables it contains are randomised over, which is the case sometimes.} that, when given a state $\sigma$ and a deterministic expression $e$, returns this single value.

The definition is obvious and leads to the following equality (assuming that $e$ is deterministic and well-typed and $\sigma$ is a valid state):
$$\mathrm{expr\_sem}\ \sigma\ e\ =\ \mathrm{return}\ (\mathrm{expr\_sem\_rf}\ \sigma\ e)$$

This property will later enable us to also convert deterministic source-language expressions into \enquote{equivalent} target-language expressions.

\begin{figure}[!hb]
\begin{oframed}
\vspace*{-6mm}
\begin{align*}
&\mathbf{primrec}\ \mathrm{expr\_sem}\ ::\ \mathrm{state}\Rightarrow\mathrm{expr}\Rightarrow\mathrm{val\ measure}\ \mathbf{where}\\[-1mm]
&\hskip1.3mm\ \mathrm{expr\_sem}\ \sigma\ (\mathrm{Val}\ v)\ =\ \mathrm{return\_val}\ v\\[-1mm]
&|\ \mathrm{expr\_sem}\ \sigma\ (\mathrm{Var}\ x)\ =\ \mathrm{return\_val}\ (\sigma\ x)\\[-1mm]
&|\ \mathrm{expr\_sem}\ \sigma\ (\mathrm{LET}\ e_1\ \mathrm{IN}\ e_2)\ =\ \\[-1mm]
&\hskip12mm \mathbf{do}\ \{\\[-1mm]
&\hskip17mm v\leftarrow \mathrm{expr\_sem}\ \sigma\ e_1;\\[-1mm]
&\hskip17mm \mathrm{expr\_sem}\ (v\dbshift\sigma)\ e_2\\[-1mm]
&\hskip12mm \}\\[-1mm]
&|\ \mathrm{expr\_sem}\ \sigma\ (\pdfop{\mathit{op}} e)\ =\ \\[-1mm]
&\hskip12mm \mathbf{do}\ \{\\[-1mm]
&\hskip17mm v\leftarrow \mathrm{expr\_sem}\ \sigma\ e;\\[-1mm]
&\hskip17mm \mathrm{return\_val}\ (\mathrm{op\_sem}\ \mathit{op}\ v)\\[-1mm]
&\hskip12mm \}\\[-1mm]
&|\ \mathrm{expr\_sem}\ \sigma\ \pdfpair{e_1}{e_2}\ =\ \\[-1mm]
&\hskip12mm \mathbf{do}\ \{\\[-1mm]
&\hskip17mm v\leftarrow \mathrm{expr\_sem}\ \sigma\ e_1;\\[-1mm]
&\hskip17mm w\leftarrow \mathrm{expr\_sem}\ \sigma\ e_2;\\[-1mm]
&\hskip17mm \mathrm{return\_val}\ \pdfvalpair{v}{w}\\[-1mm]
&\hskip12mm \}\\[-1mm]
&|\ \mathrm{expr\_sem}\ \sigma\ (\mathrm{IF}\ b\ \mathrm{THEN}\ e_1\ \mathrm{ELSE}\ e_2)\ =\ \\[-1mm]
&\hskip12mm \mathbf{do}\ \{\\[-1mm]
&\hskip17mm b'\leftarrow \mathrm{expr\_sem}\ \sigma\ b;\\[-1mm]
&\hskip17mm \mathbf{if}\ b' = \mathrm{TRUE}\ \mathbf{then}\ \mathrm{expr\_sem}\ \sigma\ e_1\ \mathbf{else}\ \mathrm{expr\_sem}\ \sigma\ e_2\\[-1mm]
&\hskip12mm \}\\[-1mm]
&|\ \mathrm{expr\_sem}\ \sigma\ (\mathrm{Random}\ \mathit{dst}\ e)\ =\ \\[-1mm]
&\hskip12mm \mathbf{do}\ \{\\[-1mm]
&\hskip17mm p\leftarrow \mathrm{expr\_sem}\ \sigma\ e;\\[-1mm]
&\hskip17mm \mathrm{dist\_measure}\ \mathit{dst}\ p\\[-1mm]
&\hskip12mm \}\\[-1mm]
&|\ \mathrm{expr\_sem}\ \sigma\ (\mathrm{Fail}\ t)\ =\ \mathrm{null\_measure}\ (\mathrm{stock\_measure}\ t)
\end{align*}
\vspace*{-8mm}
\end{oframed}
\caption{The semantics of source-language expressions}
\label{expr_sem}
\end{figure}

\subsection{Target language}
\label{sec_target}

The target language is again modelled very closely after the one used by Bhat\etAl~\cite{bhat13}. 
The type system and the operators are the same as in the source language, but the key difference is that while 
expressions in the source language return a measure space on their result type, the expressions in the target language 
always return a single value.

Since our source language lacks sum types, so does our target language. Additionally, our target language 
differs from that used by Bhat\etAl in the following respects:
\begin{itemize}
\item Our language has no function types; since functions only occur as integrands and as final results (as the compilation result is a density function), we can simply define integration to introduce the integration variable as a bound variable and let the final result contain a single free variable with de~Bruijn index $0$, \ie there is an implicit $\uplambda$ abstraction around the compilation result.
\item Evaluation of expressions in our target language can never fail. In the language by Bhat\etAl, failure is used to handle undefined integrals; we, on the other hand, use the convention of Isabelle's measure theory library, which returns 0 for integrals of non-integrable functions. This has the advantage of keeping the semantics simple, which makes proofs considerably easier.
\item Our target language does not have Let bindings, since, in contrast to the source language, they would be semantically superfluous here. However, they are still useful in practice since they yield shorter expressions and can avoid multiple evaluation of the same term; they could be added with little effort.
\end{itemize}

Figures \ref{target_lang}, \ref{cexpr_typing}, and \ref{cexpr_sem} show the syntax, typing rules, and semantics of the target language.

\begin{figure}[p]
\begin{oframed}
\vspace*{-4mm}
\begin{align*}
&\mathbf{datatype}\ \mathrm{cexpr}\ =\ \\
&\hskip10mm\mathrm{CVar}\ \mathit{nat}\ |\ \mathrm{CVal}\ \mathit{val}\ |\ \pdfcop{\mathit{pdf\hskip-0.4mm\_operator}} \mathit{cexpr}\ |\ \hskip-0.9mm\pdfcpair{\mathit{cexpr}}{\mathit{cexpr}}\hskip-1mm\ |\ \\
&\hskip10mm\mathrm{IF}_\mathrm{c}\ \mathit{cexpr}\ \mathrm{THEN}\ \mathit{cexpr}\ \mathrm{ELSE}\ \mathit{cexpr}\ |\ \pdfint{\mathit{cexpr}}{\mathit{pdf\hskip-0.5mm\_type}}
\end{align*}
\vspace*{-6mm}
\end{oframed}
\caption{The expressions of the target language}
\label{target_lang}
\end{figure}

\begin{figure}[p]
\begin{oframed}
\begin{center}
\begin{tabular}{ccc}
$\inferrule[cet\_val]{ }{\cexprtyping{\Gamma}{\mathrm{CVal}\ v}{\mathrm{val\_type}\ v}}$ & \quad\quad &
$\inferrule[cet\_var]{ }{\cexprtyping{\Gamma}{\mathrm{CVar}\ x}{\Gamma\ x}}$\\[8mm]
$\inferrule[cet\_op]{\cexprtyping{\Gamma}{e}{t} \\ \mathrm{op\_type}\ \mathit{op}\ t = \mathrm{Some}\ t'}
          {\cexprtyping{\Gamma}{\,\pdfcop{\textit{op}} e}{t'}}$ & &
$\inferrule[cet\_pair]{\cexprtyping{\Gamma}{e_1}{t_1} \\ \cexprtyping{\Gamma}{e_2}{t_2}}
          {\cexprtyping{\Gamma}{\,\pdfcpair{e_1}{e_2}\,}{t_1\hskip-0.6mm \times t_2}}$\\[8mm]
$\inferrule[cet\_if]{\cexprtyping{\Gamma}{b}{\BB} \\ \cexprtyping{\Gamma}{e_1}{t} \\ \cexprtyping{\Gamma}{e_2}{t}}
          {\cexprtyping{\Gamma}{\mathrm{IF}_\mathrm{c}\ b\ \mathrm{THEN}\ e_1\ \mathrm{ELSE}\ e_2}{t}}$ & &
$\inferrule[cet\_int]{\cexprtyping{t\dbshift\Gamma}{e}{\RR}}
          {\cexprtyping{\Gamma}{\hskip-0.2mm\pdfint{\hskip0.3mm e\hskip0.2mm}{t}}{\RR}}$
\end{tabular}
\end{center}
\vspace*{-4mm}
\end{oframed}
\caption{The typing rules for the target language}
\label{cexpr_typing}
\end{figure}

\begin{figure}[p]
\begin{oframed}
\vspace*{-4mm}
\begin{align*}
&\mathbf{primrec}\ \mathrm{cexpr\_sem}\ ::\ \mathrm{state}\Rightarrow\mathrm{cexpr}\Rightarrow\mathrm{val}\ \mathbf{where}\\
&\hskip1.3mm\ \mathrm{cexpr\_sem}\ \sigma\ (\mathrm{CVal}\ v)\ =\ v\\
&|\ \mathrm{cexpr\_sem}\ \sigma\ (\mathrm{CVar}\ x)\ =\ \sigma\ x\\
&|\ \mathrm{cexpr\_sem}\ \sigma\ \pdfcpair{e_1}{e_2}\ =\ \pdfvalpair{\mathrm{cexpr\_sem}\ \sigma\ e_1}{\mathrm{cexpr\_sem}\ \sigma\ e_2}\\
&|\ \mathrm{cexpr\_sem}\ \sigma\ \pdfcop{\mathit{op}} e\ =\ \mathrm{op\_sem}\ \mathit{op}\ (\mathrm{cexpr\_sem}\ \sigma\ e)\\
&|\ \mathrm{cexpr\_sem}\ \sigma\ (\mathrm{IF}_\mathrm{c}\ b\ \mathrm{THEN}\ e_1\ \mathrm{ELSE}\ e_2)\ =\\
&\hskip12mm(\mathbf{if}\ \mathrm{cexpr\_sem}\ \sigma\ b = \mathrm{TRUE}\ \mathbf{then}\ \mathrm{cexpr\_sem}\ \sigma\ e_1\ \mathbf{else}\ \mathrm{cexpr\_sem}\ \sigma\ e_2)\\
&|\ \mathrm{cexpr\_sem}\ \sigma\ (\pdfint{e}{t})\ =\\
&\hskip12mm\mathrm{RealVal}\ ({\textstyle\int}\hskip-0.3mm x.\ \mathrm{extract\_real}\ (\mathrm{cexpr\_sem}\ (x\dbshift\sigma)\ e)\,\dd\hskip0.3mm\mathrm{stock\_measure}\ t)
\end{align*}
\vspace*{-6mm}
\end{oframed}
\caption{The semantics of the target language}
\label{cexpr_sem}
\end{figure}

\paragraph{Converting deterministic expressions.}
The auxiliary function $\mathit{expr\_rf\_to\_cexpr}$, which will be used in some rules of the compiler that handle deterministic expressions, is of particular interest. We mentioned earlier that deterministic source-language expressions can be converted to equivalent target-language expressions.\footnote{Bhat\etAl say that a deterministic expression \enquote{is also an expression in the target language syntax, and we silently treat it as such}~\cite{bhat13}} This function does precisely that. Its definition is obvious.

$\mathit{expr\_rf\_to\_cexpr}$ satisfies the following equality for any deterministic source-language expression $e$:
$$\mathrm{cexpr\_sem}\ \sigma\ (\mathrm{expr\_rf\_to\_cexpr}\ e)\ =\ \mathrm{expr\_sem\_rf}\ \sigma\ e$$
\

\section{Abstract compiler}
\label{sec_abstract}

\subsection{Density contexts}

First, we define the notion of a \emph{density context}, which holds the acquired context data the compiler will require to compute 
the density of an expression. A density context is a tuple $\Upsilon = (V, V', \Gamma, \delta)$ that contains the following information:
\begin{itemize}
\item The set $V$ of random variables in the current context. These are the variables that are randomised over.
\item The set $V'$ of parameter variables in the current context. These are variables that may occur in the expression, 
but are not randomised over but treated as constants.
\item The type environment $\Gamma$
\item A density function $\delta$ that returns the common density of the variables $V$ under the parameters $V'$. Here, $\delta$ 
is a function from $\mathit{space}\ (\mathit{state\_measure}\ (V\cup V')\ \Gamma)$ to the extended real numbers.
\end{itemize}

A density context $(V,V',\Gamma,\delta)$ describes a parametrised measure on the states on $V\cup V'$. 
Let $\rho \in \mathit{space}\ (\mathit{state\_measure}\ V'\ \Gamma)$ be a set of parameters. Then we write
$\mathit{dens\_ctxt\_measure}\ (V,V',\Gamma,\delta)\ \rho$ \noindent for the measure that we obtain by taking $\mathit{state\_measure}\ V\ \Gamma$, transforming it by merging a given state $\sigma$ with the parameter state $\rho$ and finally applying the density $\delta$ on the resulting image measure. The Isabelle definition of this is:
{\small
\begin{align*}
&\mathbf{definition}\ \mathrm{dens\_ctxt\_measure}\ ::\ \mathrm{dens\_ctxt}\Rightarrow\mathrm{state}\Rightarrow\mathrm{state\_measure}\ \mathbf{where}\\
&\hskip6mm \mathrm{dens\_ctxt\_measure}\ (V,V',\Gamma,\delta)\ \rho\ =\ \mathrm{density}\ (\mathrm{distr}\ (\mathrm{state\_measure}\ V\ \Gamma)\\
&\hskip20mm (\mathrm{state\_measure}\ (V\cup V')\ \Gamma)\ \,(\uplambda\sigma.\ \mathrm{merge}\ V\ V'\ (\sigma, \rho)))\ \delta
\end{align*}}

Informally, $\mathit{dens\_ctxt\_measure}$ describes the measure obtained by integrating over the variables $v_1,\ldots,v_m\in V$ while treating the variables $v_1',\ldots, v_n'\in V'$ as parameters. The evaluation of an expression $e$ with variables from $V\cup V'$ in this context is effectively a function
\begin{align*}
&\uplambda v_1'\,\ldots\,v_n'.\ \posint v_1.\ \ldots\posint v_m.\ \,\mathrm{expr\_sem}\ (v_1,\ldots,v_m,v_1',\ldots,v_n')\ e\ \cdot\\
&\hskip26mm \delta\ (v_1,\ldots,v_m,v_1',\ldots,v_n')\ \dd \Gamma\ v_1\ldots\dd \Gamma\ v_m\;.
\end{align*}

\noindent A density context is \emph{well-formed} (implemented in Isabelle as the locale $\mathit{density\_context}$) if:
\begin{itemize}
\item $V$ and $V'$ are finite and disjoint
\item $\delta\ \sigma \geq 0$ for any $\sigma \in \mathrm{space}\ (\mathrm{state\_measure}\ (V\cup V')\ \Gamma)$
\item $\delta$ is Borel-measurable \wrt $\mathrm{state\_measure}\ (V\cup V')\ \Gamma$
\item for any $\rho\in\mathrm{space}\ (\mathrm{state\_measure}\ V'\ \Gamma)$, the $\mathrm{dens\_ctxt\_measure}\ (V,V',\Gamma,\delta)\ \rho$ is a sub-probability measure
\end{itemize}

\subsection{Definition}
\label{abstract_compiler}

As a first step, we have implemented an abstract density compiler as an inductive predicate $\exprcompd{\Upsilon}{e}{f}$, 
where $\Upsilon$ is a density context, $e$ is a source-language expression and $f$ is a function of type 
$\mathrm{val}\ \mathrm{state} \Rightarrow \mathrm{val} \Rightarrow \mathrm{ereal}$\;. Its first parameter is a state that assigns values to the free variables 
in $e$ and its second parameter is the value for which the density is to be computed. The compiler therefore computes 
a density function that is parametrised with the values of the non-random free variables in the source expression.

The compilation rules are virtually identical to those by Bhat\etAl~\cite{bhat13}, except for the following adaptations:
\begin{itemize}
\item Bhat\etAl handle the $\mathbf{IF}$\,\dots\,$\mathbf{THEN}$\,\dots\,$\mathbf{ELSE}$ case with the \enquote{match} rule for sum types. As we do not 
support sum types, we have a dedicated rule for $\mathbf{IF}$\,\dots\,$\mathbf{THEN}$\, \dots\,$\mathbf{ELSE}\,\dots$\;.
\item The use of de~Bruijn indices requires shifting of variable sets and states whenever the scope of a new bound variable is entered; 
unfortunately, this makes some rules somewhat technical.
\item We do not provide any compiler support for \emph{deterministic} \enquote{let} bindings. They are semantically redundant, as they can always be expanded without changing the semantics of the expression. In fact, they \emph{have} to be unfolded for compilation, so they can be regarded as a feature that adds convenience, but no expressivity.
\end{itemize}

\ 

The following list shows the standard compilation rules adapted from Bhat\etAl, plus a rule for multiplication with a constant.\footnote{Additionally, three congruence rules are required for technical reasons. These rules are required because the abstract and the concrete result may differ on a null set and outside their domain.} Note that the functions $\mathit{marg\_dens}$ and $\mathit{marg\_dens2}$ compute the marginal density of one (resp. two) variables by \enquote{integrating away} all the other variables from the common density $\delta$. The function $\mathit{branch\_prob}$ computes the probability of being in the current branch of execution by integrating over \emph{all} the variables in the common density $\delta$.
\begin{oframed}
\small
\noindent\textsc{hd\_val}\vskip-4mm
$$\inferrule{\mathrm{countable\_type}\ (\mathrm{val\_type}\ v)}
{\exprcompd{(V,V',\Gamma,\delta)}{\mathrm{Val}\ v}{\uplambda\rho\, x.\ \mathrm{branch\_prob}\ (V,V'\Gamma,\delta)\ \rho \cdot \langle x = v\rangle}}$$
\vskip2mm

\noindent\textsc{hd\_var}\vskip-4mm
$$\inferrule{x \in V}{\exprcompd{(V,V',\Gamma,\delta)}{\mathrm{Var}\ x}{\mathrm{marg\_dens}\ (V,V',\Gamma,\delta)\ x}}$$
\vskip2mm

\noindent\textsc{hd\_pair}\vskip-4mm
$$\inferrule{x \in V \\ y \in V \\ x \neq y}{\exprcompd{(V,V',\Gamma,\delta)}{\pdfpair{\mathrm{Var}\ x}{\mathrm{Var}\ y}}{\mathrm{marg\_dens2}\ (V,V',\Gamma,\delta)\ x\ y}}$$
\vskip2mm

\noindent\textsc{hd\_fail}\vskip-4mm   
$$\inferrule{ }{\exprcompd{(V,V',\Gamma,\delta)}{\mathrm{Fail}\ t}{\uplambda\rho\,x.\ 0}}$$
\vskip2mm
          
\noindent\textsc{hd\_let}\nopagebreak\vskip-4mm\nopagebreak
$$\inferrule{\exprcompd{(\emptyset ,V\cup V',\Gamma ,\uplambda x.\ 1)}{e_1}{f} \\
\exprcompd{(0\dbshift V, \{x+1\ |\ x\in V'\}, \mathrm{type\_of}\ \Gamma\ e_1\dbshift\Gamma, (f\cdot\delta)(\rho \circ (\uplambda x.\ x+1))}{e_2}{g}}
{\exprcompd{(V,V',\Gamma,\delta)}{\mathrm{LET}\ e_1\ \mathrm{IN}\ e_2}{\uplambda\rho.\ g\ (\mathrm{undefined}\dbshift\rho)}}$$
\vskip2mm
          
\noindent\textsc{hd\_rand}\nopagebreak\vskip-4mm\nopagebreak
$$\mprset{flushleft}\inferrule{\exprcompd{(V,V',\Gamma,\delta)}{e}{f}}{\exprcompd{(V,V',\Gamma,\delta)}
        {\mathrm{Random}\ \mathit{dst}\ e}{}\\\\
        \hspace*{10mm}\uplambda\rho\,y.\ {\textstyle\int^{+}}\hskip-1mm x.\ f\ \rho\ x\cdot\mathrm{dist\_dens}\ \mathit{dst}\ x\ y\ \dd\hskip0.3mm \mathrm{dist\_param\_type}\ \mathit{dst}}$$
\vskip2mm

\noindent\textsc{hd\_rand\_det}\nopagebreak\vskip-4mm\nopagebreak
$$\inferrule{e\ \mathrm{det} \\ \mathrm{free\_vars}\ e \subseteq V'}
            {\exprcompd{(V,V',\Gamma,\delta)}{\mathrm{Random}\ \mathit{dst}\ e}{\hskip70mm}\\\\ \uplambda \rho\,x.\ \mathrm{branch\_prob}\ (V,V',\Gamma,\delta)\ \rho\cdot \mathrm{dist\_dens}\ \mathit{dst}\ (\mathrm{expr\_sem\_rf}\ \rho\ e)\ x}$$
\vskip2mm

\noindent\textsc{hd\_if}\vskip-4mm
$$\inferrule{\exprcompd{(\emptyset, V\cup V',\Gamma,\uplambda\rho.\ 1)}{b}{f} \\
             \exprcompd{(V,V',\Gamma,\uplambda\rho.\ \delta\ \rho \cdot f\ \mathrm{TRUE})}{e_1}{g_1} \\
             \exprcompd{(V,V',\Gamma,\uplambda\rho.\ \delta\ \rho \cdot f\ \mathrm{FALSE})}{e_2}{g_2}}
            {\exprcompd{(V,V',\Gamma,\delta)}{\mathrm{IF}\ b\ \mathrm{THEN}\ e_1\ \mathrm{ELSE}\ e_2}{\uplambda\rho\,x.\ g_1\ \rho\ x + g_2\ \rho\ x}}$$
\vskip2mm
            
\noindent\textsc{hd\_if\_det}\vskip-4mm
$$\inferrule{b\ \mathrm{det} \\\\
        \exprcompd{(V,V',\Gamma, \uplambda\rho.\ \delta\ \rho\cdot \langle\mathrm{expr\_sem\_rf}\ \rho\ b = \mathrm{TRUE}\rangle)}{e_1}{g_1} \\
        \exprcompd{(V,V',\Gamma, \uplambda\rho.\ \delta\ \rho\cdot \langle\mathrm{expr\_sem\_rf}\ \rho\ b = \mathrm{FALSE}\rangle)}{e_2}{g_2}}
        {\exprcompd{(V,V',\Gamma,\delta)}{\mathrm{IF}\ b\ \mathrm{THEN}\ e_1\ \mathrm{ELSE}\ e_2}{\uplambda\rho\,x.\ g_1\ \rho\ x + g_2\ \rho\ x}}$$
\vskip2mm
        
\noindent\textsc{hd\_fst}\vskip-4mm
$$\inferrule{\exprcompd{(V,V',\Gamma,\delta)}{e}{f}}{\exprcompd{(V,V',\Gamma,\delta)}{\,\mathrm{fst}\ e}{\uplambda\rho\,x.\ {\textstyle\int^+}\hskip-1mm y.\ f\ \rho\ \pdfvalpair{x}{y}\ \dd\hskip0.3mm\mathrm{type\_of}\ \Gamma\ (\mathrm{snd}\ e)}}$$
\vskip2mm

\noindent\textsc{hd\_snd}\vskip-4mm
$$\inferrule{\exprcompd{(V,V',\Gamma,\delta)}{e}{f}}{\exprcompd{(V,V',\Gamma,\delta)}{\,\mathrm{snd}\ e}{\uplambda\rho\,y.\ {\textstyle\int^+}\hskip-1mm x.\ f\ \rho\ \pdfvalpair{x}{y}\ \dd\hskip0.3mm\mathrm{type\_of}\ \Gamma\ (\mathrm{fst}\ e)}}$$
\vskip2mm

\noindent\textsc{hd\_op\_discr}\vskip-4mm
$$\inferrule{\mathrm{countable\_type}\ (\mathrm{type\_of}\ (\pdfop{\mathit{op}} e)) \\ \exprcompd{(V,V',\Gamma,\delta)}{e}{f}}
            {\exprcompd{(V,V',\Gamma,\delta)}{\,\pdfop{\mathit{op}} e}{\uplambda\rho\,y.\ {\textstyle\int^+}\hskip-1mm x.\ \langle\mathrm{op\_sem}\ \mathit{op}\ x = y\rangle\cdot f\ \rho\ x\ \dd\hskip0.3mm\mathrm{type\_of}\ \Gamma\ e}}$$
\vskip2mm
            
\noindent\textsc{hd\_neg}\vskip-4mm
$$\inferrule{\exprcompd{(V,V',\Gamma,\delta)}{e}{f}}{\exprcompd{(V,V',\Gamma,\delta)}{-e}{\uplambda\rho\,x.\ f\ \rho\ \,(-x)}}$$
\vskip2mm

\noindent\textsc{hd\_addc}\vskip-4mm
$$\inferrule{e'\ \mathrm{det} \\ \mathrm{free\_vars}\ e' \subseteq V' \\ \exprcompd{(V,V',\Gamma,\delta)}{e}{f}}
            {\exprcompd{(V,V',\Gamma,\delta)}{e + e'}{\uplambda\rho\,x.\ f\ \rho\ \,(x - \mathrm{expr\_sem\_rf}\ \rho\ e')}}$$
\vskip2mm

\noindent\textsc{hd\_multc}\vskip-4mm
$$\inferrule{c \neq 0 \\ \exprcompd{(V,V',\Gamma,\delta)}{e}{f}}
            {\exprcompd{(V,V',\Gamma,\delta)}{e \cdot \mathrm{Val}\ (\mathrm{RealVal}\ c)}{\uplambda\rho\,x.\ f\ \rho\ \,(x / c)\, /\, |c|}}$$
\vskip2mm
            
\noindent\textsc{hd\_add}\vskip-4mm
$$\inferrule{\exprcompd{(V,V',\Gamma,\delta)}{\pdfpair{e_1}{e_2}}{f}}
            {\exprcompd{(V,V',\Gamma,\delta)}{e_1 + e_2}{\uplambda\rho\,z.\ {\textstyle\int^+}\hskip-1mm x.\ f\ \rho\,\pdfvalpair{x}{z - x}\,\dd\hskip0.3mm\mathrm{type\_of}\ \Gamma\ e_1}}$$
\vskip2mm

\noindent\textsc{hd\_inv}\nopagebreak\vskip-4mm\nopagebreak
$$\inferrule{\exprcompd{(V,V',\Gamma,\delta)}{e}{f}}
            {\exprcompd{(V,V',\Gamma,\delta)}{e^{-1}}{\uplambda\rho\,x.\ f\ \rho\ (x^{-1})\,/\,x^2}}$$
\vskip2mm
            
\noindent\textsc{hd\_exp}\nopagebreak\vskip-4mm\nopagebreak
$$\inferrule{\exprcompd{(V,V',\Gamma,\delta)}{e}{f}}
            {\exprcompd{(V,V',\Gamma,\delta)}{\mathrm{exp}\ e}{\uplambda\rho\,x.\ \mathbf{if}\ x > 0\ \mathbf{then}\ f\ \rho\ \,(\ln x)\,/\,x\ \mathbf{else}\ 0}}$$
\vskip2mm            
\end{oframed}
\begin{figure}[!hp]
\vspace*{-6mm}
\caption{The abstract compilation rules}
\label{expr_has_density}
\end{figure}

\subsection{Soundness proof}

We show the following soundness result for the abstract compiler:
\footnote{Note that since the abstract compiler returns parametrised density functions, we need to parametrise the 
result with the state $\uplambda x.\ \mathrm{undefined}$, even if the expression contains no free variables.}
\begin{oframed}
\vspace*{-5mm}
\small
\begin{align*}
&\hspace*{-1.5mm}\textbf{lemma}\ \mathrm{expr\_has\_density\_sound}:\\[-0.75mm]
&\hskip2.5mm\mathbf{assumes}\ \ \exprcompd{(\emptyset,\emptyset,\Gamma,\uplambda\rho.\ 1)}{e}{f}\hskip2mm\mathbf{and}\hskip2mm\exprtyping{\Gamma}{e}{t}\hskip2mm\mathbf{and}\hskip3mm\mathrm{free\_vars}\ e\ =\ \emptyset\\[-0.75mm]
&\hskip2.5mm\mathbf{shows}\ \mathrm{has\_subprob\_density}\ (\mathrm{expr\_sem}\ \sigma\ e)\ (\mathrm{stock\_measure}\ t)\ (f\ (\uplambda x.\ \mathrm{undefined}))\hspace*{-2mm}
\end{align*}
\vspace*{-7mm}
\end{oframed}

\noindent Here, $\mathit{has\_subprob\_density}\ M\ N\ f$ is an abbreviation for the following four facts:
\begin{itemize}
\item applying the density $f$ to $N$ yields $M$
\item $M$ is a sub-probability measure
\item $f$ is $N$-Borel-measurable
\item $f$ is non-negative on its domain
\end{itemize}

\noindent We prove this with the following generalised auxiliary lemma:
\begin{oframed}
\vspace*{-5mm}
\small
\begin{align*}
&\hspace*{-2.5mm}\textbf{lemma}\ \mathrm{expr\_has\_density\_sound\_aux}:\\[-0.75mm]
&\hskip2mm\mathbf{assumes}\ \ \exprcompd{(V,V',\Gamma,\delta)}{e}{f}\hskip3mm\mathbf{and}\hskip3mm\exprtyping{\Gamma}{e}{t}\hskip3mm\mathbf{and}\hskip3mm\mathrm{free\_vars}\ e = \emptyset\\[-0.75mm]
&\hskip21.6mm \mathbf{and}\hskip3mm\mathrm{density\_context}\ V\ V'\ \Gamma\ \delta\hskip3mm\mathbf{and}\hskip3mm\mathrm{free\_vars}\ e\ \subseteq\ V\cup V'\\[-0.75mm]
&\hskip2mm\mathbf{shows}\hskip6.0mm\mathrm{has\_parametrized\_subprob\_density}\ (\mathrm{state\_measure}\ V'\ \Gamma)\\[-0.75mm]
&\hskip23.6mm (\uplambda\rho.\ \mathbf{do}\ \{\sigma\leftarrow\mathrm{dens\_ctxt\_measure}\ (V,V',\Gamma,\delta)\ \rho;\ \mathrm{expr\_sem}\ \sigma\ e\})\hskip-2mm\\[-0.75mm]
&\hskip23.6mm (\mathrm{stock\_measure\ t})\ f
\end{align*}
\vspace*{-7mm}
\end{oframed}

The predicate $\mathit{has\_parametrized\_subprob\_density}\ R\ M\ N\ f$ simply means that $f$ is Borel-measurable \wrt $R\otimes N$ and 
that for any parameter state $\rho$ from $R$, the predicate $\mathit{has\_subprob\_density}\ M\ N\ f$ holds.\\

The proof is by straightforward induction following the inductive definition of the abstract compiler. In many cases, the monad laws 
for the Giry monad allow restructuring the induction goal in such a way that the induction hypothesis can be applied directly; 
in the other cases, the definitions of the monadic operations need to be unfolded and the goal is essentially to show that two integrals are equal and that the output produced is well-formed.

The proof given by Bhat\etAl~\cite{bhat_journal} is analogous to ours, but much more concise due to the fact that side conditions such as measurability, integrability, non-negativity, and so on are not proven explicitly and many important (but uninteresting) steps are skipped or only hinted at.

\section{Concrete compiler}
\label{sec_concrete}

\subsection{Approach}

The concrete compiler is another inductive predicate, modelled directly after the abstract compiler, but returning a target-language expression as the compilation result instead of a HOL function. We will use a standard refinement approach to relate the concrete compiler to the abstract compiler. We thus lift the soundness result on the abstract compiler to an analogous one on the concrete one. This effectively shows that the concrete compiler always returns a well-formed target-language expression that represents a density for the sub-probability space described by the source language.

The concrete compilation predicate is written as $$\exprcompc{(\mathit{vs}, \mathit{vs}', \Gamma, \delta)}{e}{f}$$
\indent Here, $\mathit{vs}$ and $\mathit{vs}'$ are lists of variables, $\Gamma$ is a typing environment, and $\delta$ is a target-language expression describing the common density of the random variables $\mathit{vs}$ in the context. It may be parametrised with the variables from $\mathit{vs}'$.

\subsection{Definition}
\label{concrete_compiler}

The concrete compilation rules are, of course, a direct copy of the abstract ones, but with all the abstract HOL operations replaced with operations on target-language expressions. Due to the de~Bruijn indices and the lack of functions as explicit objects in the target language, some of the rules are somewhat complicated -- inserting an expression into the scope of one or more bound variables (such as under an integral) requires shifting the variable indices of the inserted expression correctly. For this reason, we will not print the rules here; they can be found in the Isabelle theory file \texttt{PDF\_Compiler.thy}.

\subsection{Refinement}

The refinement relates the concrete compilation $\exprcompc{(\mathit{vs}, \mathit{vs}', \Gamma, \delta)}{e}{f}$ to the abstract compilation
$$\exprcompc{(\text{set}\ vs, \text{set}\ vs', \Gamma, \uplambda\sigma.\ \text{cexpr\_sem}\ \sigma\ \delta)}{e}{\uplambda \rho\,x.\ \text{cexpr\_sem}\ (x\dbshift\rho)\ f}$$

\noindent In words: we take the abstract compilation predicate and
\begin{itemize}
\item variable sets are refined to variable lists
\item the typing context and the source-language expression remain unchanged
\item the common density in the context and the compilation result are refined from HOL functions to target-language expressions (by applying the target language semantics)
\end{itemize}

The main refinement lemma states that the concrete compiler yields a result that is equivalent to that of the abstract compiler, modulo refinement. Informally, the statement is the following: if $e$ is ground and well-typed under some well-formed concrete density context $\Upsilon$ and $\exprcompc{\Upsilon}{e}{f}$, then $\exprcompd{\Upsilon'}{e}{f'}$, where $\Upsilon'$ and $f'$ are the abstract versions of $\Upsilon$ and $f$.

The proof for this is conceptually simple -- induction over the definition of the concrete compiler; in practice, however, it is quite involved. In every single induction step, the well-formedness of the intermediary expressions needs to be shown, congruence lemmas for the abstract compiler need to be applied, and, when integration is involved, non-negativity and integrability have to be shown in order to convert non-negative integrals to Lebesgue integrals and integrals on product spaces to iterated integrals.

Combining this main refinement lemma and the abstract soundness lemma, we can now easily show the concrete soundness lemma:
\begin{oframed}
\vspace*{-6mm}
\begin{align*}
&\textbf{lemma}\ \mathrm{expr\_has\_density\_cexpr\_sound}:\\[-0.75mm]
&\hskip5mm\mathbf{assumes}\ \ \exprcompc{([],[],\Gamma,1)}{e}{f}\hskip3mm\mathbf{and}\hskip3mm\exprtyping{\Gamma}{e}{t}\hskip3mm\mathbf{and}\hskip3mm\mathrm{free\_vars}\ e\ =\ \emptyset\\[-0.75mm]
&\hskip5mm\mathbf{shows}\ \hskip5mm\mathrm{has\_subprob\_density}\ (\mathrm{expr\_sem}\ \sigma\ e)\ (\mathrm{stock\_measure}\ t)\\[-0.75mm]
&\hskip53.6mm
(\uplambda x.\ \text{cexpr\_sem}\ (x\dbshift\sigma)\ f)\\[-0.75mm]
&\hskip21.32mm \Gamma'\ 0 = t \Longrightarrow \cexprtyping{\Gamma'}{f}{\text{REAL}}\\[-0.75mm]
&\hskip21.32mm \text{free\_vars}\ f \subseteq \{0\}
\end{align*}
\vspace*{-8mm}
\end{oframed}
\indent Informally, the lemma states that if $e$ is a well-typed, ground source-language expression, compiling it with an empty context will yield a well-typed, well-formed target-language expression representing a density function on the measure space described by $e$.

\subsection{Final result}
\label{compiles_to}
We will now summarise the soundness lemma we have just proven in a more concise manner.
For convenience, we define the symbol $\exprcomp{e}{t}{f}$ (read \enquote{$e$ with type $t$ compiles to $f$}), which includes 
the well-typedness and groundness requirements on $e$ as well as the compilation result:\footnote{In this definition, the choice of the typing environment is, of course, completely arbitrary since the expression contains no free variables.}\vspace*{-1mm}
$$\exprcomp{e}{t}{f}\hskip2mm \longleftrightarrow\hskip2mm (\exprtyping{\uplambda x.\ \mathrm{UNIT}}{e}{t}\hskip1mm \wedge\hskip1mm \mathrm{free\_vars}\ e = \emptyset\hskip1mm \wedge\hskip1mm
([], [], \uplambda x.\ \mathrm{UNIT}, 1) \turnstile_\mathrm{c} e \Rightarrow f)$$

\noindent The final soundness theorem for the compiler, stated in Isabelle syntax, is then:\footnote{To be precise, the lemma statement in Isabelle is slightly different; for better readability, 
we unfolded one auxiliary definition here and omitted the type cast from real to ereal.}
\begin{oframed}
\vspace*{-6mm}
\begin{align*}
&\mathbf{lemma}\ \mathrm{expr\_compiles\_to\_sound}:\\[-1mm]
&\hskip0mm\mathbf{assumes}\ \ e : t \Rightarrow_\mathrm{c}\hskip-0.5mm f\\[-1mm]
&\hskip0mm\mathbf{fixes}\ \Gamma\ \Gamma'\ \sigma\ \sigma'\\[-1mm]
&\hskip0mm\mathbf{shows}\hskip2.0mm \mathrm{expr\_sem}\ \sigma\ e = \mathrm{density}\ (\mathrm{stock\_measure}\ t)\ (\uplambda x.\ \mathrm{cexpr\_sem}\ (x\dbshift\sigma')\ f)\hspace*{-2mm}\\[-1mm]
&\hskip12.8mm \forall x\hskip-0.5mm\in\hskip-0.4mm\mathrm{type\_universe}\ t.\ \mathrm{cexpr\_sem}\ (x\dbshift\sigma')\ f \geq 0\\[-1mm]
&\hskip12.8mm \exprtyping{\Gamma}{e}{t}\\[-1mm]
&\hskip12.8mm \cexprtyping{t\dbshift\Gamma'}{f}{\mathrm{REAL}}\\[-1mm]
&\hskip12.8mm \mathrm{free\_vars}\ f \subseteq \{0\}
\end{align*}
\vspace*{-8mm}
\end{oframed}

\noindent In words, this result means the following:
\begin{center}
\begin{oframed}
\raggedright
\noindent\textbf{Theorem}\vskip2mm
\hrule\vskip3mm
\noindent Let $e$ be a source-language expression. If the compiler determines that $e$ is well-formed and well-typed with type $t$ 
and returns the target-language expression $f$, then:\vspace*{-1mm}
\begin{itemize}
\item the measure obtained by taking the stock measure of $t$ and using the evaluation of $f$ as a density is the measure obtained by evaluating $e$
\item $f$ is non-negative on all input values of type $t$
\item $e$ has no free variables and indeed has type $t$ (in any type context $\Gamma$)
\item $f$ has no free variable except the parameter (\ie the variable 0) and is a function from $t$ to $\mathit{REAL}$\footnote{meaning if its parameter variable has type $t$, it is of type REAL}
\end{itemize}
\end{oframed}
\end{center}

Isabelle's code generator now allows us to execute our inductively-defined verified compiler using the \textbf{values} command\footnote{Our compiler is inherently non-deterministic since it may return zero, one, or many density functions, seeing as an expression may have no matching compilation rules or more than one. Therefore, we must use the \textbf{values} command instead of the \textbf{value} command and receive a set of compilation results.} or generate code in one of the target languages such as Standard ML or Haskell.

\subsection{Evaluation}

As an example on which to test the compiler, we choose the same expression that was chosen by Bhat\etAl~\cite{bhat13}:\footnote{$\mathrm{Val}$ and $\mathrm{RealVal}$ were omitted for better readability and symbolic variable names were used instead of de~Bruijn indices.}
\begin{align*}
&\mathbf{LET}\ x = \mathrm{Random\ UniformReal}\,\pdfpair{0}{1}\,\mathbf{IN}\\
&\hskip5mm\mathbf{LET}\ y = \mathrm{Random\ Bernoulli}\ x\ \mathbf{IN}\\
&\hskip10mm\mathbf{IF}\ y\ \mathbf{THEN}\ x\ +\ 1\ \mathbf{ELSE}\ x
\end{align*}

We abbreviate this expression with $e$. We can then display the result of the compilation using the Isabelle 
command $\mathbf{values}\ "\{(t, f)\ |\,t\ f.\ \,\exprcomp{e}{t}{f}\}"\;$.

The result is a singleton set which contains the pair $(\mathrm{REAL}, f)$, where $f$ is a very long and complicated 
expression. Simplifying constant subexpressions and expressions of the form $\mathrm{fst}\,\pdfpair{e_1}{e_2}$ and again 
replacing de~Bruijn indices with symbolic identifiers, we obtain:
{\small
\begin{align*}
&{\textstyle\int} b.\ (\mathbf{IF}\ 0 \leq x-1 \wedge x-1 \leq 1\ \mathbf{THEN}\ 1\ \mathbf{ELSE}\ 0) \cdot (\mathbf{IF}\ 0 \leq x-1 \wedge x-1 \leq 1\ \mathbf{THEN}\\
&\hskip15mm \mathbf{IF}\ b\ \mathbf{THEN}\ x-1\ \mathbf{ELSE}\ 1-(x-1)\ \mathbf{ELSE}\ 0)\cdot\langle b\rangle\ +\\
&{\textstyle\int} b.\ (\mathbf{IF}\ 0 \leq x \wedge x \leq 1\ \mathbf{THEN}\ 1\ \mathbf{ELSE}\ 0) \cdot (\mathbf{IF}\ 0 \leq x \wedge x \leq 1\ \mathbf{THEN}\\
&\hskip15mm \mathbf{IF}\ b\ \mathbf{THEN}\ x\ \mathbf{ELSE}\ 1-x\ \mathbf{ELSE}\ 0)\cdot\langle \neg\, b\rangle
\end{align*}}

\noindent Further simplification yields the following result:
$$\langle 1\leq x\leq 2\rangle\cdot(x-1) + \langle 0\leq x\leq 1\rangle\cdot (1-x)$$

While this result is the same as that which Bhat\etAl have reached, our compiler generates a much larger expression than the one they printed. The reason for this is that they printed a $\upbeta$-reduced version of the compiler output; in particular, constant subexpressions were evaluated. While such simplification is, of course, very useful when using the compiler in practice, we have not implemented it since it is not conceptually interesting and outside the scope of this work.

\section{Conclusion}
\label{sec_conclusion}

\subsection{Breakdown}

All in all, the formalisation of the compiler took about three months. It contains a total of roughly 10000 lines of Isabelle code (definitions, lemma statements, proofs, and examples). Table \ref{breakdown} shows a detailed breakdown of this.
\begin{table}[!htp]
\caption{Breakdown of the Isabelle code}
\begin{center}
\begin{tabular}{ll}
\toprule
Type system and semantics & $2900$ lines\\
Abstract compiler & $2600$ lines\\
Concrete compiler & $1400$ lines\\
General Measure Theory / auxiliary lemmas \quad\quad & $3400$ lines\\
\bottomrule
\end{tabular}
\end{center}
\label{breakdown}
\vspace{-6mm}
\end{table}

As can be seen from this table, a sizeable portion of the work was the formalisation of results from general measure theory, such as the substitution lemmas for Lebesgue integrals, and auxiliary notions and lemmas, such as measure embeddings. Since the utility of these formalisations is not restricted to this particular project, they will be moved to Isabelle's Measure Theory library.

\subsection{Difficulties}

The main problems we encountered during the formalisation were:
\paragraph{Missing background theory.} As mentioned in the previous section, a sizeable amount of Measure Theory and auxiliary notions had to be formalised. Most notably, the existing Measure Theory library did not contain integration by substitution. As a side product of their formalisation of the Central Limit Theorem~\cite{avigad}, Avigad\etAl proved the Fundamental Theorem of Calculus and, building thereupon, integration by substitution. However, their integration-by-substitution lemma only supported continuous functions, whereas we required the theorem for general Borel-measurable functions. Using their proof of the Fundamental Theorem of Calculus, we proved such a lemma, which initially comprised almost 1000 lines of proof, but has been shortened significantly thereafter.

\paragraph{Proving side conditions.} Many lemmas from the Measure Theory library require measurability, integrability, non-negativity, etc. In hand-written proofs, this is often \enquote{hand-waved} or implicitly dismissed as trivial; in a formal proof, proving these can blow up proofs and render them very complicated and technical. The measurability proofs in particular are ubiquitous in our formalisation. The Measure Theory library provides some tools for proving measurability automatically, but while they were quite helpful in many cases, they are still work in progress and require more tuning.

\paragraph{Lambda calculus.} Bhat\etAl use a simply-typed Lambda-calculus-like language with symbolic identifiers as a target language. For a paper proof, this is the obvious choice, since it leads to concise and familiar definitions, but in a formal setting, it always comes with the typical problems of having to deal with variable capture and related issues. For that reason, we chose to use de~Bruijn indices instead; however, this makes handling target-language terms less intuitive, since variable indices need to be shifted whenever several target-language terms are combined.

Another issue was the lack of a function type in our target language; allowing first-order functions, as Bhat\etAl did, would have made many definitions easier and more natural, but would have complicated others significantly due to measurability issues.

Furthermore, we effectively needed to formalise a number of properties of Lambda calculus that can be used implicitly in a paper proof.

\subsection{Future work}

The following improvements to the Isabelle formalisation could probably be realised with little effort:
\begin{itemize}
\item sum types and a \textbf{match}\,\ldots\,\textbf{with}\,\ldots statement
\item a compiler rule for deterministic \enquote{let} bindings
\item \enquote{let} bindings for the target language
\item a preprocessing stage to allow \enquote{normal} variables instead of de~Bruijn indices
\item a postprocessing stage to automatically simplify the density expression as far as possible
\end{itemize}

The first of these is interesting because it is the only substantial weakness of our formalisation as opposed to that by Bhat\etAl; the remaining four merely make using the compiler more convenient or more efficient.

Additionally, in the long term, a \emph{Markov-chain Monte Carlo (MCMC)} sampling method such as the \emph{Metropolis--Hastings algorithm} could also be formalised in Isabelle and integrated with the density compiler. However, this would be a more involved project as it will require the formalisation of additional probability theory in Isabelle.

\subsection{Summary}

Using Isabelle\slash HOL, we formalised the semantics of a simple probabilistic functional programming language with predefined probability distributions and a compiler that returns the probability distribution that a program in this language describes. These are modelled very closely after those given by Bhat\etAl~\cite{bhat13}. We then used the existing formalisations of measure theory in Isabelle\slash HOL to formally prove the correctness of this compiler \wrt the semantics of the source and target languages.

This shows not only that the compiler given by Bhat\etAl is correct, but also that a formal correctness proof for such a compiler can be done with reasonable effort and that Isabelle\slash HOL in general and its Measure Theory library in particular are suitable for it.

{
\raggedright

}

\newpage
\begin{appendix}

\noindent{\LARGE\bfseries Appendix}
\renewcommand{\arraystretch}{1.4}

\section{Operator semantics}
\renewcommand{\arraystretch}{1.5}
\begin{center}
\small
\begin{longtable}{cccc}
\toprule
\quad\textsc{Operator}\quad & \quad\textsc{Input type}\quad & \quad\textsc{Output type}\quad & \quad\textsc{Semantics}\quad\\\midrule
\multirow{2}{*}{Add}   & $\ZZ\times\ZZ$   & $\ZZ$   & \multirow{2}{*}{$a + b$}\\[-0.5mm]
                       & $\RR\times\RR$     & $\RR$    & \\\midrule
\multirow{2}{*}{Minus} & $\ZZ$                       & $\ZZ$   & \multirow{2}{*}{$-a$}\\[-0.5mm]
                       & $\RR$                        & $\RR$    & \\\midrule
\multirow{2}{*}{Mult}  & $\ZZ\times\ZZ$   & $\ZZ$   & \multirow{2}{*}{$a \cdot b$}\\[-0.5mm]
                       & $\RR\times\RR$     & $\RR$    & \\\midrule
Inverse                & $\RR$                        & $\RR$    & \rule{0pt}{4.3ex} $\begin{cases}\frac{1}{a}& \mathrm{for}\ a \neq 0\\0 & \mathrm{otherwise}\end{cases}$ \rule[-3.1ex]{0pt}{3.1ex}\\[3mm]\midrule
Sqrt                   & $\RR$                        & $\RR$    & \rule{0pt}{4.3ex} $\begin{cases}\sqrt{a}& \mathrm{for}\ a \geq 0\\0 & \mathrm{otherwise}\end{cases}$ \rule[-3.1ex]{0pt}{3.1ex} \\[2mm]\midrule
Exp                    & $\RR$                        & $\RR$    & $e^a$\\[0.7mm]\midrule
Ln                     & $\RR$                        & $\RR$    & \rule{0pt}{4.3ex} $\begin{cases}\ln a & \mathrm{for}\ a > 0\\0 & \mathrm{otherwise}\end{cases}$ \rule[-3.1ex]{0pt}{3.1ex} \\[2mm]\midrule
Pi                     & $\mathrm{UNIT}$                        & $\RR$    & $\uppi$\\\midrule
\multirow{2}{*}{Pow}   &  $\ZZ\times\ZZ$  & $\ZZ$   & \multirow{2}{*}{$ \begin{cases}a^b & \mathrm{for}\ a \neq 0, b \geq 0\\ 0 & \mathrm{otherwise}\end{cases}$}\\[-0.5mm]
                       & $\RR\times\ZZ$  & $\RR$    & \\\midrule
Fact                   & $\ZZ$                       & $\ZZ$   & $a!$\\[0.7mm]\midrule
And                    & $\BB\times\BB$     & $\BB$    & $a \wedge b$\\\midrule
Or                     & $\BB\times\BB$     & $\BB$    & $a \vee b$\\\midrule
Not                    & $\BB$                        & $\BB$    & $\neg\, a$\\\midrule
Equals                 & $t \times t$                           & $\BB$    & $a = b$\\\midrule
\multirow{2}{*}{Less}  & $\ZZ\times\ZZ$   & $\BB$    & \multirow{2}{*}{$a < b$}\\[-0.5mm]
                       & $\RR\times\RR$     & $\BB$    & \\\midrule
Fst                    & $t_1 \times t_2$                       & $t_1$              & $a$ for $z=(a,b)$\\[0.5mm]\midrule
Snd                    & $t_1 \times t_2$                       & $t_2$              & $b$ for $z=(a,b)$\\[0.5mm]\midrule
\multirow{2}{*}{Cast REAL} & $\BB$                        & $\RR$    & $\langle a=\mathrm{TRUE}\rangle$ \\[-0.5mm]
                       & $\ZZ$                       & $\RR$    & $a$ as a real number\\[1mm]\midrule

\multirow{2}{*}{Cast INT} & $\BB$                        & $\ZZ$   & $\langle a=\mathrm{TRUE}\rangle$ \\[-0.5mm]
                       & $\RR$                        & $\ZZ$   & $\lfloor a\rfloor$\\[1mm]
\bottomrule
\end{longtable}
\end{center}
\renewcommand{\arraystretch}{1.4}

\section{Built-in distributions}

\begin{figure}[h!]
\footnotesize
\renewcommand{\arraystretch}{1.3}
\begin{center}
\begin{tabular}{cccccc}
\toprule
\hspace*{1mm}\textsc{Distribution}\hspace*{1mm} & \hspace*{1mm}\textsc{Param.}\hspace*{1mm} & \hspace*{1mm}\textsc{Domain}\hspace*{1mm} & \hspace*{2mm} & \textsc{Type} &  \textsc{Density function}\\
\midrule
Bernoulli & $\RR$ & $p\in[0;1]$ && $\BB$ & \rule{0pt}{4.3ex} $\begin{cases} p & \mathrm{for}\ x=\mathrm{TRUE}\\ 1-p&\mathrm{for}\ x=\mathrm{FALSE}\end{cases}$ \rule[-3.1ex]{0pt}{3.1ex}\\[6mm]
UniformInt & $\ZZ\times\ZZ$ & $p_1\leq p_2$ && $\ZZ$ & $\dfrac{\langle x \in [p_1;p_2]\rangle}{p_2-p_1+1}$\\[6mm]
UniformReal & $\RR\times\RR$ & $p_1 < p_2$ && $\RR$ & $\dfrac{\langle x \in [p_1;p_2]\rangle}{p_2-p_1}$\\[4mm]
Gaussian & $\RR\times\RR$ & $p_2>0$ && $\RR$ & \rule{0pt}{4.6ex} $\dfrac{1}{\sqrt{2\uppi p_2^2}}\ \,\mathrm{exp}\left(-\dfrac{(x-p_1)^2}{2p_2^2}\right)$ \rule[-3.4ex]{0pt}{3.4ex}\\[8mm]
Poisson & $\RR$ & $p\geq 0$ && $\ZZ$ & \rule{0pt}{4.3ex} $\begin{cases} \mathrm{exp}(-p)\cdot p^x / x! & \mathrm{for\ }x\geq 0\\ 0 & \mathrm{otherwise}\end{cases}$ \rule[-3.1ex]{0pt}{3.1ex}\\[3mm]
\bottomrule
\end{tabular}
\end{center}
\renewcommand{\arraystretch}{1}

\caption{The built-in distributions of the source language.\\
\footnotesize
The density functions are given in terms of the parameter $p$, which is of the type given in the column \enquote{parameter type}. If $p$ is of a product type, $p_1$ and $p_2$ stand for the two components of $p$. $x$ is the function variable, \eg the point at which the density function is to be evaluated.}
\label{pdf_dists}
\end{figure}

\newpage

\section{Target language auxiliary functions}

For the definitions of the following functions, see the corresponding Isabelle theory file \texttt{PDF\_Target\_Semantics.thy}. We will not print them here since they are rather technical.

In this section, $x$ will always be a variable, $v$ will always be a value, and $e$ and $e'$ will always be target language expressions. $f$ and $g$ will always be target language expressions interpreted as functions, \ie with an implicit $\uplambda$ abstraction around them.

All functions in the table take de Bruijn indices into account, \ie they will shift variables when they enter the scope of an integral.\\

\label{appendix_target_aux}
{\small
\noindent\begin{tabularx}{\textwidth}{lX}
\toprule
\textsc{Function} & \textsc{Description} \\\midrule
$\textrm{case\_nat}\ y\ f\ x$ & $\mathbf{if}\ x = 0\ \mathbf{then}\ y\ \mathbf{else}\ f\ (x - 1)$\\
$\textrm{map\_vars}\ h\ e$ & transforms all variables in $e$ with the function $h$\\
$\textrm{ins\_var0}\ e$ & $\mathrm{map\_vars}\ \mathrm{Suc}\ e$\newline (prepares expression $e$ for insertion into scope of new variable)\\
$\textrm{ins\_var1}\ f$ & $\mathrm{map\_vars}\ (\mathrm{case\_nat}\ 0\ (\uplambda x.\ x+2))\ f$\newline (prepares function $f$ for insertion into scope of new variable)\\
$\textrm{del\_var0}\ e$ & $\mathrm{map\_vars}\ (\uplambda x.\ x-1)\ e$\newline (shifts all variables down, deleting the variable $0$)\\
$\textrm{cexpr\_subst}\ x\ e\ e'$ & substitutes $e$ for $x$ in $e'$\\
$\textrm{cexpr\_subst\_val}\ e\ v$ & substitutes $v$ for $\mathrm{CVar}\ 0$ in $e$, effectively applying $e$ interpreted as a function to the argument $v$.\\
$\textrm{cexpr\_comp}\ f\ g$ & function composition, \ie $\uplambda x.\ f\ (g\ x)$\\
$f\circ_c g$ & alias for $\mathrm{cexpr\_comp}$\\
$\textrm{expr\_rf\_to\_cexpr}\ e$ & converts a deterministic source language expression into a target language expression, see Sect. \ref{sec_target}\\
$\textrm{integrate\_var}\ \Gamma\ x\ e$ & integrates $e$ over the free variable $x$\\
$\textrm{integrate\_vars}\ \Gamma\ \mathit{xs}\ e$ & integrates $e$ over the free variables in the list $\mathit{xs}$\\
$\textrm{branch\_prob\_cexpr}\ (\mathit{vs}, \mathit{vs}', \Gamma, \delta)$ \hspace*{2mm} & returns an expression that computes $\mathrm{branch\_prob}$\\
$\textrm{marg\_dens\_cexpr}\ \Gamma\ \mathit{vs}\ x\ \delta$  & returns an expression that computes the $\mathrm{marg\_dens}$\\
$\textrm{marg\_dens2\_cexpr}\ \Gamma\ \mathit{vs}\ x\ y\ \delta$  & returns an expression that computes the $\mathrm{marg\_dens2}$\\
$\textrm{dist\_dens\_cexpr}\ \mathit{dst}\ e\ e'$ & returns an expression that computes the density of the built-in distribution $\mathit{dst}$, parametrised with $e$ and evaluated at $e'$\\
\bottomrule
\end{tabularx}
}

\newpage

\section{Concrete compiler rules}

\begin{oframed}
\footnotesize
\noindent{\normalsize\textsc{edc\_val}}\vskip-6mm
$$\mprset{flushleft}\inferrule{\mathrm{countable\_type}\ (\mathrm{val\_type}\ v)}
{\exprcompc{(\mathit{vs}, \mathit{vs}', \Gamma, \delta)}{\mathrm{Val}\ v}{}\\\\
\hspace*{10mm}  {\mathrm{ins\_var0}\ (\mathrm{branch\_prob\_cexpr}\ (\mathit{vs}, \mathit{vs}', \Gamma, \delta)) 
         \cdot \langle\mathrm{CVar}\ 0 = \mathrm{CVal}\ v\rangle}}$$
\vskip2mm
         
\noindent{\normalsize\textsc{edc\_var}}\vskip-4mm
$$\inferrule{x \in \mathrm{set}\ \mathit{vs}}
{\exprcompc{(\mathit{vs}, \mathit{vs}', \Gamma, \delta)}{\mathrm{Var}\ x}
  {\mathrm{marg\_dens\_cexpr}\ \Gamma\ \mathit{vs}\ x\ \delta}}$$
\vskip2mm
  
\noindent{\normalsize\textsc{edc\_pair}}\vskip-4mm
$$\inferrule{x \in \mathrm{set}\ \mathit{vs} \\ y \in \mathrm{set}\ \mathit{vs}}
{\exprcompc{(\mathit{vs}, \mathit{vs}', \Gamma, \delta)}{\pdfpair{x}{y}}
  {\mathrm{marg\_dens2\_cexpr}\ \Gamma\ \mathrm{vs}\ x\ y\ \delta}}$$  
\vskip2mm

\noindent{\normalsize\textsc{edc\_fail}}\nopagebreak\vskip-4mm\nopagebreak
$$\inferrule{ }{\exprcompc{(\mathit{vs}, \mathit{vs}', \Gamma, \delta)}{\mathrm{Fail}\ t}{\mathrm{CVal}\ (\mathrm{RealVal}\ 0)}}$$  
\vskip2mm

\noindent{\normalsize\textsc{edc\_let}}\nopagebreak\vskip-6mm\nopagebreak
$$\inferrule{\exprcompc{([], \mathit{vs}\mathbin{@}\mathit{vs}', \Gamma, 1)}{e}{f} \\
             \exprcompc{(0\,\#\,\mathrm{map}\ \mathrm{Suc}\ \mathit{vs},\ \mathrm{map}\ \mathrm{Suc}\ \mathit{vs}',\ \mathrm{type\_of}\ \Gamma\ e\,\dbshift\,\Gamma,\ \mathrm{ins\_var0}\ \delta\,\cdot\, f)}{e'}{g}}
    {\exprcompc{(\mathit{vs}, \mathit{vs}', \Gamma, \delta)}{\mathrm{LET}\ e\ \mathrm{IN}\ e'}{\mathrm{del\_var0}\ g}}$$  
\vskip2mm

\noindent{\normalsize\textsc{edc\_rand}}\nopagebreak\vskip-4mm\nopagebreak
$$\mprset{flushleft}\inferrule{\exprcompc{(vs, \mathit{vs}', \Gamma, \delta)}{e}{f}}
    {\exprcompc{(\mathit{vs}, \mathit{vs}', \Gamma, \delta)}{\mathrm{Random}\ \mathit{dst}\ e}{}\\\\
    \hspace*{10mm}{\textstyle\int_c}\ \mathrm{ins\_var1}\ f\ \cdot\ \mathrm{dist\_dens\_cexpr}\ \mathit{dst}\ (\mathrm{CVar}\ 0)\ (\mathrm{CVar}\ 1)\\\\\hspace*{20mm} {\dd\hskip0.3mm\mathrm{dist\_param\_type}\ \mathit{dst}}}$$
\vskip2mm

\noindent{\normalsize\textsc{edc\_rand\_det}}\nopagebreak\vskip-6mm\nopagebreak
$$\mprset{flushleft}\inferrule{e\ \mathrm{det} \\ \mathrm{free\_vars}\ e \subseteq \mathrm{set}\ \mathit{vs}'}
    {\exprcompc{(\mathit{vs}, \mathit{vs}', \Gamma, \delta)}{\mathrm{Random}\ \mathit{dst}\ e}{}\\\\
    \hspace*{10mm}\mathrm{ins\_var0}\ (\mathrm{branch\_prob\_cexpr}\ (\mathit{vs}, \mathit{vs}', \Gamma, \delta))\ \cdot \\\\
    \hspace*{10mm}    \mathrm{dist\_dens\_cexpr}\ \mathit{dst}\ (\mathrm{ins\_var0}\ (\mathrm{expr\_rf\_to\_cexpr\ e}))\ (\mathrm{CVar}\ 0)}$$
\vskip2mm

\noindent{\normalsize\textsc{edc\_if}}\vskip-4mm
$$\inferrule{\exprcompc{([], \mathit{vs}\mathbin{@}\mathit{vs}', \Gamma, 1)}{b}{f} \\\\
             \exprcompc{(\mathit{vs}, \mathit{vs}', \Gamma, \delta \cdot \langle\mathrm{cexpr\_subst\_val}\ f\ \mathrm{TRUE}\rangle)}{e_1}{f_1} \\\\
             \exprcompc{(\mathit{vs}, \mathit{vs}', \Gamma, \delta \cdot \langle\mathrm{cexpr\_subst\_val}\ f\ \mathrm{FALSE}\rangle)}{e_2}{f_2}}
    {\exprcompc{(\mathit{vs}, \mathit{vs}', \Gamma, \delta)}{\mathrm{IF}\ b\ \mathrm{THEN}\ e_1\ \mathrm{ELSE}\ e_2}{f_1 + f_2}}$$
\vskip2mm

\noindent{\normalsize\textsc{edc\_if\_det}}\vskip-6mm
$$\inferrule{b\ \mathrm{det} \\\\
             \exprcompc{(\mathit{vs}, \mathit{vs}', \Gamma, \delta \cdot \langle\mathrm{expr\_rf\_to\_cexpr}\ b\rangle)}{e_1}{f_1} \\\\
             \exprcompc{(\mathit{vs}, \mathit{vs}', \Gamma, \delta \cdot \langle\mathrm{\neg\ expr\_rf\_to\_cexpr}\ b\rangle)}{e_2}{f_2}}
    {\exprcompc{(\mathit{vs}, \mathit{vs}', \Gamma, \delta)}{\mathrm{IF}\ b\ \mathrm{THEN}\ e_1\ \mathrm{ELSE}\ e_2}{f_1 + f_2}}$$
\vskip2mm

\noindent{\normalsize\textsc{edc\_op\_discr}}\nopagebreak\vskip-6mm\nopagebreak
$$\inferrule{\exprtyping{\Gamma}{e}{t} \\ \mathrm{op\_type}\ \mathit{oper}\ t = \mathrm{Some}\ t' \\ \mathrm{countable\_type}\ t' \\
             \exprcompc{(\mathit{vs}, \mathit{vs}', \Gamma, \delta)}{e}{f}}
    {\exprcompc{(\mathit{vs}, \mathit{vs}', \Gamma, \delta)}{\pdfop{\mathit{oper}} e}
     {{\textstyle\int_c}\ \langle \pdfcop{\mathit{oper}} \mathrm{CVar}\ 0 = \mathrm{CVar}\ 1\rangle \cdot 
          \mathrm{ins\_var1}\ f\ \dd t}}$$
\vskip2mm

\noindent{\normalsize\textsc{edc\_fst}}\vskip-4mm
$$\inferrule{\exprtyping{\Gamma}{e}{t\times t'} \\
             \exprcompc{(\mathit{vs}, \mathit{vs}', \Gamma, \delta)}{e}{f}}
    {\exprcompc{(\mathit{vs}, \mathit{vs}', \Gamma, \delta)}{\mathrm{fst}\ e}
     {{\textstyle\int_c}\ \mathrm{ins\_var1}\ f\ \circ_c\pdfcpair{\mathrm{CVar}\ 1}{\mathrm{CVar}\ 0} \dd t'}}$$
\vskip2mm

\noindent{\normalsize\textsc{edc\_snd}}\vskip-4mm
$$\inferrule{\exprtyping{\Gamma}{e}{t\times t'} \\
             \exprcompc{(\mathit{vs}, \mathit{vs}', \Gamma, \delta)}{e}{f}}
    {\exprcompc{(\mathit{vs}, \mathit{vs}', \Gamma, \delta)}{\mathrm{snd}\ e}
     {{\textstyle\int_c}\ \mathrm{ins\_var1}\ f\ \circ_c\pdfcpair{\mathrm{CVar}\ 0}{\mathrm{CVar}\ 1} \dd t}}$$
\vskip2mm

\noindent{\normalsize\textsc{edc\_neg}}\nopagebreak\vskip-4mm\nopagebreak
$$\inferrule{\exprcompc{(\mathit{vs}, \mathit{vs}', \Gamma, \delta)}{e}{f}}
    {\exprcompc{(\mathit{vs}, \mathit{vs}', \Gamma, \delta)}{-e}{f\circ_c(-\mathrm{CVar}\ 0)}}$$
\vskip2mm

\noindent{\normalsize\textsc{edc\_addc}}\nopagebreak[4]\vskip-5mm
\nopagebreak[4]
$$\inferrule{\exprcompc{(\mathit{vs}, \mathit{vs}', \Gamma, \delta)}{e}{f} \\ e'\ \mathrm{det} \\ \mathrm{free\_vars}\ e' \subseteq \mathrm{set}\ \mathit{vs}'}
    {\exprcompc{(\mathit{vs}, \mathit{vs}', \Gamma, \delta)}{e + e'}{f\circ_c(\mathrm{CVar}\ 0 - \mathrm{ins\_var0}\ (\mathrm{expr\_rf\_to\_cexpr}\ e'))}}$$
\vskip2mm

\noindent{\normalsize\textsc{edc\_multc}}\nopagebreak\vskip-4mm\nopagebreak
$$\inferrule{\exprcompc{(\mathit{vs}, \mathit{vs}', \Gamma, \delta)}{e}{f} \\ c \neq 0}
    {\exprcompc{(\mathit{vs}, \mathit{vs}', \Gamma, \delta)}{e \cdot \mathrm{Val}\ (\mathrm{RealVal}\ c)}
      {(f\circ_c (\mathrm{CVar}\ 0\,/\, c)) / |c|}}$$
\vskip2mm

\noindent{\normalsize\textsc{edc\_add}}\nopagebreak\vskip-4mm\nopagebreak
$$\inferrule{\exprtyping{\Gamma}{\pdfpair{e}{e'}}{t\times t} \\ \exprcompc{(\mathit{vs}, \mathit{vs}', \Gamma, \delta)}{\pdfpair{e}{e'}}{f}}
    {\exprcompc{(\mathit{vs}, \mathit{vs}', \Gamma, \delta)}{e + e'}
        {{\textstyle\int_c}\  \mathrm{ins\_var1}\ f\ \circ_c\pdfpair{\mathrm{CVar}\ 0}{\mathrm{CVar}\ 1 - \mathrm{CVar}\ 0\,} \dd t}}$$
\vskip2mm

\noindent{\normalsize\textsc{edc\_inv}}\nopagebreak\vskip-4mm\nopagebreak
$$\inferrule{\exprcompc{(\mathit{vs}, \mathit{vs}', \Gamma, \delta)}{e}{f}}
    {\exprcompc{(\mathit{vs}, \mathit{vs}', \Gamma, \delta)}{e^{-1}\hskip-1mm}
        {(f \circ_c (1 / \mathrm{CVar}\ 0))\, /\, (\mathrm{CVar}\ 0)^2}}$$
\vskip2mm

\noindent{\normalsize\textsc{edc\_exp}}\vskip-6mm
$$\mprset{flushleft}\inferrule{\exprcompc{(\mathit{vs}, \mathit{vs}', \Gamma, \delta)}{e}{f}}
    {\exprcompc{(\mathit{vs}, \mathit{vs}', \Gamma, \delta)}{\mathrm{exp}\ e}{}\\\\
        \hspace*{10mm}{\mathrm{IF}\ \,\mathrm{CVar}\ 0 > 0\,\ \mathrm{THEN}\ (f\circ_c \mathrm{ln}\ (\mathrm{CVar}\ 0))\,/\,\mathrm{CVar}\ 0\ \mathrm{ELSE}\ 0}}$$
\end{oframed}

\end{appendix}


\begin{thebibliography}{10}
\providecommand{\url}[1]{\texttt{#1}}
\providecommand{\urlprefix}{URL }

\bibitem{audebaud}
Audebaud, P., Paulin-Mohring, C.: Proofs of randomized algorithms in {C}oq. In:
  Mathematics of Program Construction, Lecture Notes in Computer Science, vol.
  4014, pp. 49--68. Springer Berlin Heidelberg (2006),
  \url{http://dx.doi.org/10.1007/11783596_6}

\bibitem{avigad}
Avigad, J., H{\"o}lzl, J., Serafin, L.: A formally verified proof of the
  {C}entral {L}imit {T}heorem. CoRR  abs/1405.7012 (2014)

\bibitem{bhat12}
Bhat, S., Agarwal, A., Vuduc, R., Gray, A.: A type theory for probability
  density functions. In: Proceedings of the 39th Annual ACM SIGPLAN-SIGACT
  Symposium on Principles of Programming Languages. pp. 545--556. POPL '12,
  ACM, New York, NY, USA (2012),
  \url{http://doi.acm.org/10.1145/2103656.2103721}

\bibitem{bhat13}
Bhat, S., Borgstr\"om, J., Gordon, A.D., Russo, C.: Deriving probability
  density functions from probabilistic functional programs. In: Tools and
  Algorithms for the Construction and Analysis of Systems, Lecture Notes in
  Computer Science, vol. 7795, pp. 508--522. Springer Berlin Heidelberg (2013),
  \url{http://dx.doi.org/10.1007/978-3-642-36742-7_35}, (best paper award)

\bibitem{bhat_journal}
Bhat, S., Borgstr\"om, J., Gordon, A.D., Russo, C.: Deriving probability
  density functions from probabilistic functional programs. In: TODO. Springer
  Berlin Heidelberg (2014)

\bibitem{cock}
Cock, D.: Verifying probabilistic correctness in {I}sabelle with {pGCL}. In:
  Proceedings of the 7th Systems Software Verification. pp. 1--10 (November
  2012)

\bibitem{cock_isa}
Cock, D.: {pGCL} for {I}sabelle. Archive of Formal Proofs  (Jul 2014),
  \url{http://afp.sf.net/entries/pGCL.shtml}, Formal proof development

\bibitem{doberkat_book}
Doberkat, E.E.: Stochastic relations: foundations for {M}arkov transition
  systems. Studies in {I}nformatics, Chapman \& Hall/CRC (2007)

\bibitem{doberkat}
Doberkat, E.E.: Basing {M}arkov transition systems on the {G}iry monad.
  \url{http://www.informatics.sussex.ac.uk/events/domains9/Slides/Doberkat_GiryMonad.pdf}
  (2008)

\bibitem{giry}
Giry, M.: A categorical approach to probability theory. In: Categorical Aspects
  of Topology and Analysis, Lecture Notes in Mathematics, vol. 915, pp. 68--85.
  Springer Berlin Heidelberg (1982), \url{http://dx.doi.org/10.1007/BFb0092872}

\bibitem{hoelzl}
H\"olzl, J.: Construction and stochastic applications of measure spaces in
  {H}igher-{O}rder {L}ogic. PhD thesis, Technische Universit\"at M\"unchen,
  Institut f\"ur Informatik (2012)

\bibitem{hurd}
Hurd, J., McIver, A., Morgan, C.: Probabilistic guarded commands mechanized in
  {HOL}. Electron. Notes Theor. Comput. Sci.  112,  95--111 (Jan 2005),
  \url{http://dx.doi.org/10.1016/j.entcs.2004.01.021}

\bibitem{park05}
Park, S., Pfenning, F., Thrun, S.: A probabilistic language based upon sampling
  functions. In: Proceedings of the 32Nd ACM SIGPLAN-SIGACT Symposium on
  Principles of Programming Languages. pp. 171--182. POPL '05, ACM, New York,
  NY, USA (2005), \url{http://doi.acm.org/10.1145/1040305.1040320}

\end{thebibliography}
\end{document}